\documentclass[useAMS,usegraphicx,usenatbib]{mn2e} 
\usepackage{ulem}
\usepackage{color}
\definecolor{AliceBlue}{rgb}{0.94,0.97,1.00}
\definecolor{AntiqueWhite1}{rgb}{1.00,0.94,0.86}
\definecolor{AntiqueWhite2}{rgb}{0.93,0.87,0.80}
\definecolor{AntiqueWhite3}{rgb}{0.80,0.75,0.69}
\definecolor{AntiqueWhite4}{rgb}{0.55,0.51,0.47}
\definecolor{AntiqueWhite}{rgb}{0.98,0.92,0.84}
\definecolor{BlanchedAlmond}{rgb}{1.00,0.92,0.80}
\definecolor{BlueViolet}{rgb}{0.54,0.17,0.89}
\definecolor{CadetBlue1}{rgb}{0.60,0.96,1.00}
\definecolor{CadetBlue2}{rgb}{0.56,0.90,0.93}
\definecolor{CadetBlue3}{rgb}{0.48,0.77,0.80}
\definecolor{CadetBlue4}{rgb}{0.33,0.53,0.55}
\definecolor{CadetBlue}{rgb}{0.37,0.62,0.63}
\definecolor{CornflowerBlue}{rgb}{0.39,0.58,0.93}
\definecolor{DarkBlue}{rgb}{0.00,0.00,0.55}
\definecolor{DarkCyan}{rgb}{0.00,0.55,0.55}
\definecolor{DarkGoldenrod1}{rgb}{1.00,0.73,0.06}
\definecolor{DarkGoldenrod2}{rgb}{0.93,0.68,0.05}
\definecolor{DarkGoldenrod3}{rgb}{0.80,0.58,0.05}
\definecolor{DarkGoldenrod4}{rgb}{0.55,0.40,0.03}
\definecolor{DarkGoldenrod}{rgb}{0.72,0.53,0.04}
\definecolor{DarkGray}{rgb}{0.66,0.66,0.66}
\definecolor{DarkGreen}{rgb}{0.00,0.39,0.00}
\definecolor{DarkGrey}{rgb}{0.66,0.66,0.66}
\definecolor{DarkKhaki}{rgb}{0.74,0.72,0.42}
\definecolor{DarkMagenta}{rgb}{0.55,0.00,0.55}
\definecolor{DarkOliveGreen1}{rgb}{0.79,1.00,0.44}
\definecolor{DarkOliveGreen2}{rgb}{0.74,0.93,0.41}
\definecolor{DarkOliveGreen3}{rgb}{0.64,0.80,0.35}
\definecolor{DarkOliveGreen4}{rgb}{0.43,0.55,0.24}
\definecolor{DarkOliveGreen}{rgb}{0.33,0.42,0.18}
\definecolor{DarkOrange1}{rgb}{1.00,0.50,0.00}
\definecolor{DarkOrange2}{rgb}{0.93,0.46,0.00}
\definecolor{DarkOrange3}{rgb}{0.80,0.40,0.00}
\definecolor{DarkOrange4}{rgb}{0.55,0.27,0.00}
\definecolor{DarkOrange}{rgb}{1.00,0.55,0.00}
\definecolor{DarkOrchid1}{rgb}{0.75,0.24,1.00}
\definecolor{DarkOrchid2}{rgb}{0.70,0.23,0.93}
\definecolor{DarkOrchid3}{rgb}{0.60,0.20,0.80}
\definecolor{DarkOrchid4}{rgb}{0.41,0.13,0.55}
\definecolor{DarkOrchid}{rgb}{0.60,0.20,0.80}
\definecolor{DarkRed}{rgb}{0.55,0.00,0.00}
\definecolor{DarkSalmon}{rgb}{0.91,0.59,0.48}
\definecolor{DarkSeaGreen1}{rgb}{0.76,1.00,0.76}
\definecolor{DarkSeaGreen2}{rgb}{0.71,0.93,0.71}
\definecolor{DarkSeaGreen3}{rgb}{0.61,0.80,0.61}
\definecolor{DarkSeaGreen4}{rgb}{0.41,0.55,0.41}
\definecolor{DarkSeaGreen}{rgb}{0.56,0.74,0.56}
\definecolor{DarkSlateBlue}{rgb}{0.28,0.24,0.55}
\definecolor{DarkSlateGray1}{rgb}{0.59,1.00,1.00}
\definecolor{DarkSlateGray2}{rgb}{0.55,0.93,0.93}
\definecolor{DarkSlateGray3}{rgb}{0.47,0.80,0.80}
\definecolor{DarkSlateGray4}{rgb}{0.32,0.55,0.55}
\definecolor{DarkSlateGray}{rgb}{0.18,0.31,0.31}
\definecolor{DarkSlateGrey}{rgb}{0.18,0.31,0.31}
\definecolor{DarkTurquoise}{rgb}{0.00,0.81,0.82}
\definecolor{DarkViolet}{rgb}{0.58,0.00,0.83}
\definecolor{DeepPink1}{rgb}{1.00,0.08,0.58}
\definecolor{DeepPink2}{rgb}{0.93,0.07,0.54}
\definecolor{DeepPink3}{rgb}{0.80,0.06,0.46}
\definecolor{DeepPink4}{rgb}{0.55,0.04,0.31}
\definecolor{DeepPink}{rgb}{1.00,0.08,0.58}
\definecolor{DeepSkyBlue1}{rgb}{0.00,0.75,1.00}
\definecolor{DeepSkyBlue2}{rgb}{0.00,0.70,0.93}
\definecolor{DeepSkyBlue3}{rgb}{0.00,0.60,0.80}
\definecolor{DeepSkyBlue4}{rgb}{0.00,0.41,0.55}
\definecolor{DeepSkyBlue}{rgb}{0.00,0.75,1.00}
\definecolor{DimGray}{rgb}{0.41,0.41,0.41}
\definecolor{DimGrey}{rgb}{0.41,0.41,0.41}
\definecolor{DodgerBlue1}{rgb}{0.12,0.56,1.00}
\definecolor{DodgerBlue2}{rgb}{0.11,0.53,0.93}
\definecolor{DodgerBlue3}{rgb}{0.09,0.45,0.80}
\definecolor{DodgerBlue4}{rgb}{0.06,0.31,0.55}
\definecolor{DodgerBlue}{rgb}{0.12,0.56,1.00}
\definecolor{FloralWhite}{rgb}{1.00,0.98,0.94}
\definecolor{ForestGreen}{rgb}{0.13,0.55,0.13}
\definecolor{GhostWhite}{rgb}{0.97,0.97,1.00}
\definecolor{GreenYellow}{rgb}{0.68,1.00,0.18}
\definecolor{HotPink1}{rgb}{1.00,0.43,0.71}
\definecolor{HotPink2}{rgb}{0.93,0.42,0.65}
\definecolor{HotPink3}{rgb}{0.80,0.38,0.56}
\definecolor{HotPink4}{rgb}{0.55,0.23,0.38}
\definecolor{HotPink}{rgb}{1.00,0.41,0.71}
\definecolor{IndianRed1}{rgb}{1.00,0.42,0.42}
\definecolor{IndianRed2}{rgb}{0.93,0.39,0.39}
\definecolor{IndianRed3}{rgb}{0.80,0.33,0.33}
\definecolor{IndianRed4}{rgb}{0.55,0.23,0.23}
\definecolor{IndianRed}{rgb}{0.80,0.36,0.36}
\definecolor{LavenderBlush1}{rgb}{1.00,0.94,0.96}
\definecolor{LavenderBlush2}{rgb}{0.93,0.88,0.90}
\definecolor{LavenderBlush3}{rgb}{0.80,0.76,0.77}
\definecolor{LavenderBlush4}{rgb}{0.55,0.51,0.53}
\definecolor{LavenderBlush}{rgb}{1.00,0.94,0.96}
\definecolor{LawnGreen}{rgb}{0.49,0.99,0.00}
\definecolor{LemonChiffon1}{rgb}{1.00,0.98,0.80}
\definecolor{LemonChiffon2}{rgb}{0.93,0.91,0.75}
\definecolor{LemonChiffon3}{rgb}{0.80,0.79,0.65}
\definecolor{LemonChiffon4}{rgb}{0.55,0.54,0.44}
\definecolor{LemonChiffon}{rgb}{1.00,0.98,0.80}
\definecolor{LightBlue1}{rgb}{0.75,0.94,1.00}
\definecolor{LightBlue2}{rgb}{0.70,0.87,0.93}
\definecolor{LightBlue3}{rgb}{0.60,0.75,0.80}
\definecolor{LightBlue4}{rgb}{0.41,0.51,0.55}
\definecolor{LightBlue}{rgb}{0.68,0.85,0.90}
\definecolor{LightCoral}{rgb}{0.94,0.50,0.50}
\definecolor{LightCyan1}{rgb}{0.88,1.00,1.00}
\definecolor{LightCyan2}{rgb}{0.82,0.93,0.93}
\definecolor{LightCyan3}{rgb}{0.71,0.80,0.80}
\definecolor{LightCyan4}{rgb}{0.48,0.55,0.55}
\definecolor{LightCyan}{rgb}{0.88,1.00,1.00}
\definecolor{LightGoldenrod1}{rgb}{1.00,0.93,0.55}
\definecolor{LightGoldenrod2}{rgb}{0.93,0.86,0.51}
\definecolor{LightGoldenrod3}{rgb}{0.80,0.75,0.44}
\definecolor{LightGoldenrod4}{rgb}{0.55,0.51,0.30}
\definecolor{LightGoldenrodYellow}{rgb}{0.98,0.98,0.82}
\definecolor{LightGoldenrod}{rgb}{0.93,0.87,0.51}
\definecolor{LightGray}{rgb}{0.83,0.83,0.83}
\definecolor{LightGreen}{rgb}{0.56,0.93,0.56}
\definecolor{LightGrey}{rgb}{0.83,0.83,0.83}
\definecolor{LightPink1}{rgb}{1.00,0.68,0.73}
\definecolor{LightPink2}{rgb}{0.93,0.64,0.68}
\definecolor{LightPink3}{rgb}{0.80,0.55,0.58}
\definecolor{LightPink4}{rgb}{0.55,0.37,0.40}
\definecolor{LightPink}{rgb}{1.00,0.71,0.76}
\definecolor{LightSalmon1}{rgb}{1.00,0.63,0.48}
\definecolor{LightSalmon2}{rgb}{0.93,0.58,0.45}
\definecolor{LightSalmon3}{rgb}{0.80,0.51,0.38}
\definecolor{LightSalmon4}{rgb}{0.55,0.34,0.26}
\definecolor{LightSalmon}{rgb}{1.00,0.63,0.48}
\definecolor{LightSeaGreen}{rgb}{0.13,0.70,0.67}
\definecolor{LightSkyBlue1}{rgb}{0.69,0.89,1.00}
\definecolor{LightSkyBlue2}{rgb}{0.64,0.83,0.93}
\definecolor{LightSkyBlue3}{rgb}{0.55,0.71,0.80}
\definecolor{LightSkyBlue4}{rgb}{0.38,0.48,0.55}
\definecolor{LightSkyBlue}{rgb}{0.53,0.81,0.98}
\definecolor{LightSlateBlue}{rgb}{0.52,0.44,1.00}
\definecolor{LightSlateGray}{rgb}{0.47,0.53,0.60}
\definecolor{LightSlateGrey}{rgb}{0.47,0.53,0.60}
\definecolor{LightSteelBlue1}{rgb}{0.79,0.88,1.00}
\definecolor{LightSteelBlue2}{rgb}{0.74,0.82,0.93}
\definecolor{LightSteelBlue3}{rgb}{0.64,0.71,0.80}
\definecolor{LightSteelBlue4}{rgb}{0.43,0.48,0.55}
\definecolor{LightSteelBlue}{rgb}{0.69,0.77,0.87}
\definecolor{LightYellow1}{rgb}{1.00,1.00,0.88}
\definecolor{LightYellow2}{rgb}{0.93,0.93,0.82}
\definecolor{LightYellow3}{rgb}{0.80,0.80,0.71}
\definecolor{LightYellow4}{rgb}{0.55,0.55,0.48}
\definecolor{LightYellow}{rgb}{1.00,1.00,0.88}
\definecolor{LimeGreen}{rgb}{0.20,0.80,0.20}
\definecolor{MediumAquamarine}{rgb}{0.40,0.80,0.67}
\definecolor{MediumBlue}{rgb}{0.00,0.00,0.80}
\definecolor{MediumOrchid1}{rgb}{0.88,0.40,1.00}
\definecolor{MediumOrchid2}{rgb}{0.82,0.37,0.93}
\definecolor{MediumOrchid3}{rgb}{0.71,0.32,0.80}
\definecolor{MediumOrchid4}{rgb}{0.48,0.22,0.55}
\definecolor{MediumOrchid}{rgb}{0.73,0.33,0.83}
\definecolor{MediumPurple1}{rgb}{0.67,0.51,1.00}
\definecolor{MediumPurple2}{rgb}{0.62,0.47,0.93}
\definecolor{MediumPurple3}{rgb}{0.54,0.41,0.80}
\definecolor{MediumPurple4}{rgb}{0.36,0.28,0.55}
\definecolor{MediumPurple}{rgb}{0.58,0.44,0.86}
\definecolor{MediumSeaGreen}{rgb}{0.24,0.70,0.44}
\definecolor{MediumSlateBlue}{rgb}{0.48,0.41,0.93}
\definecolor{MediumSpringGreen}{rgb}{0.00,0.98,0.60}
\definecolor{MediumTurquoise}{rgb}{0.28,0.82,0.80}
\definecolor{MediumVioletRed}{rgb}{0.78,0.08,0.52}
\definecolor{MidnightBlue}{rgb}{0.10,0.10,0.44}
\definecolor{MintCream}{rgb}{0.96,1.00,0.98}
\definecolor{MistyRose1}{rgb}{1.00,0.89,0.88}
\definecolor{MistyRose2}{rgb}{0.93,0.84,0.82}
\definecolor{MistyRose3}{rgb}{0.80,0.72,0.71}
\definecolor{MistyRose4}{rgb}{0.55,0.49,0.48}
\definecolor{MistyRose}{rgb}{1.00,0.89,0.88}
\definecolor{NavajoWhite1}{rgb}{1.00,0.87,0.68}
\definecolor{NavajoWhite2}{rgb}{0.93,0.81,0.63}
\definecolor{NavajoWhite3}{rgb}{0.80,0.70,0.55}
\definecolor{NavajoWhite4}{rgb}{0.55,0.47,0.37}
\definecolor{NavajoWhite}{rgb}{1.00,0.87,0.68}
\definecolor{NavyBlue}{rgb}{0.00,0.00,0.50}
\definecolor{OldLace}{rgb}{0.99,0.96,0.90}
\definecolor{OliveDrab1}{rgb}{0.75,1.00,0.24}
\definecolor{OliveDrab2}{rgb}{0.70,0.93,0.23}
\definecolor{OliveDrab3}{rgb}{0.60,0.80,0.20}
\definecolor{OliveDrab4}{rgb}{0.41,0.55,0.13}
\definecolor{OliveDrab}{rgb}{0.42,0.56,0.14}
\definecolor{OrangeRed1}{rgb}{1.00,0.27,0.00}
\definecolor{OrangeRed2}{rgb}{0.93,0.25,0.00}
\definecolor{OrangeRed3}{rgb}{0.80,0.22,0.00}
\definecolor{OrangeRed4}{rgb}{0.55,0.15,0.00}
\definecolor{OrangeRed}{rgb}{1.00,0.27,0.00}
\definecolor{PaleGoldenrod}{rgb}{0.93,0.91,0.67}
\definecolor{PaleGreen1}{rgb}{0.60,1.00,0.60}
\definecolor{PaleGreen2}{rgb}{0.56,0.93,0.56}
\definecolor{PaleGreen3}{rgb}{0.49,0.80,0.49}
\definecolor{PaleGreen4}{rgb}{0.33,0.55,0.33}
\definecolor{PaleGreen}{rgb}{0.60,0.98,0.60}
\definecolor{PaleTurquoise1}{rgb}{0.73,1.00,1.00}
\definecolor{PaleTurquoise2}{rgb}{0.68,0.93,0.93}
\definecolor{PaleTurquoise3}{rgb}{0.59,0.80,0.80}
\definecolor{PaleTurquoise4}{rgb}{0.40,0.55,0.55}
\definecolor{PaleTurquoise}{rgb}{0.69,0.93,0.93}
\definecolor{PaleVioletRed1}{rgb}{1.00,0.51,0.67}
\definecolor{PaleVioletRed2}{rgb}{0.93,0.47,0.62}
\definecolor{PaleVioletRed3}{rgb}{0.80,0.41,0.54}
\definecolor{PaleVioletRed4}{rgb}{0.55,0.28,0.36}
\definecolor{PaleVioletRed}{rgb}{0.86,0.44,0.58}
\definecolor{PapayaWhip}{rgb}{1.00,0.94,0.84}
\definecolor{PeachPuff1}{rgb}{1.00,0.85,0.73}
\definecolor{PeachPuff2}{rgb}{0.93,0.80,0.68}
\definecolor{PeachPuff3}{rgb}{0.80,0.69,0.58}
\definecolor{PeachPuff4}{rgb}{0.55,0.47,0.40}
\definecolor{PeachPuff}{rgb}{1.00,0.85,0.73}
\definecolor{PowderBlue}{rgb}{0.69,0.88,0.90}
\definecolor{RosyBrown1}{rgb}{1.00,0.76,0.76}
\definecolor{RosyBrown2}{rgb}{0.93,0.71,0.71}
\definecolor{RosyBrown3}{rgb}{0.80,0.61,0.61}
\definecolor{RosyBrown4}{rgb}{0.55,0.41,0.41}
\definecolor{RosyBrown}{rgb}{0.74,0.56,0.56}
\definecolor{RoyalBlue1}{rgb}{0.28,0.46,1.00}
\definecolor{RoyalBlue2}{rgb}{0.26,0.43,0.93}
\definecolor{RoyalBlue3}{rgb}{0.23,0.37,0.80}
\definecolor{RoyalBlue4}{rgb}{0.15,0.25,0.55}
\definecolor{RoyalBlue}{rgb}{0.25,0.41,0.88}
\definecolor{SaddleBrown}{rgb}{0.55,0.27,0.07}
\definecolor{SandyBrown}{rgb}{0.96,0.64,0.38}
\definecolor{SeaGreen1}{rgb}{0.33,1.00,0.62}
\definecolor{SeaGreen2}{rgb}{0.31,0.93,0.58}
\definecolor{SeaGreen3}{rgb}{0.26,0.80,0.50}
\definecolor{SeaGreen4}{rgb}{0.18,0.55,0.34}
\definecolor{SeaGreen}{rgb}{0.18,0.55,0.34}
\definecolor{SkyBlue1}{rgb}{0.53,0.81,1.00}
\definecolor{SkyBlue2}{rgb}{0.49,0.75,0.93}
\definecolor{SkyBlue3}{rgb}{0.42,0.65,0.80}
\definecolor{SkyBlue4}{rgb}{0.29,0.44,0.55}
\definecolor{SkyBlue}{rgb}{0.53,0.81,0.92}
\definecolor{SlateBlue1}{rgb}{0.51,0.44,1.00}
\definecolor{SlateBlue2}{rgb}{0.48,0.40,0.93}
\definecolor{SlateBlue3}{rgb}{0.41,0.35,0.80}
\definecolor{SlateBlue4}{rgb}{0.28,0.24,0.55}
\definecolor{SlateBlue}{rgb}{0.42,0.35,0.80}
\definecolor{SlateGray1}{rgb}{0.78,0.89,1.00}
\definecolor{SlateGray2}{rgb}{0.73,0.83,0.93}
\definecolor{SlateGray3}{rgb}{0.62,0.71,0.80}
\definecolor{SlateGray4}{rgb}{0.42,0.48,0.55}
\definecolor{SlateGray}{rgb}{0.44,0.50,0.56}
\definecolor{SlateGrey}{rgb}{0.44,0.50,0.56}
\definecolor{SpringGreen1}{rgb}{0.00,1.00,0.50}
\definecolor{SpringGreen2}{rgb}{0.00,0.93,0.46}
\definecolor{SpringGreen3}{rgb}{0.00,0.80,0.40}
\definecolor{SpringGreen4}{rgb}{0.00,0.55,0.27}
\definecolor{SpringGreen}{rgb}{0.00,1.00,0.50}
\definecolor{SteelBlue1}{rgb}{0.39,0.72,1.00}
\definecolor{SteelBlue2}{rgb}{0.36,0.67,0.93}
\definecolor{SteelBlue3}{rgb}{0.31,0.58,0.80}
\definecolor{SteelBlue4}{rgb}{0.21,0.39,0.55}
\definecolor{SteelBlue}{rgb}{0.27,0.51,0.71}
\definecolor{VioletRed1}{rgb}{1.00,0.24,0.59}
\definecolor{VioletRed2}{rgb}{0.93,0.23,0.55}
\definecolor{VioletRed3}{rgb}{0.80,0.20,0.47}
\definecolor{VioletRed4}{rgb}{0.55,0.13,0.32}
\definecolor{VioletRed}{rgb}{0.82,0.13,0.56}
\definecolor{WhiteSmoke}{rgb}{0.96,0.96,0.96}
\definecolor{YellowGreen}{rgb}{0.60,0.80,0.20}
\definecolor{aliceblue}{rgb}{0.94,0.97,1.00}
\definecolor{antiquewhite}{rgb}{0.98,0.92,0.84}
\definecolor{aquamarine1}{rgb}{0.50,1.00,0.83}
\definecolor{aquamarine2}{rgb}{0.46,0.93,0.78}
\definecolor{aquamarine3}{rgb}{0.40,0.80,0.67}
\definecolor{aquamarine4}{rgb}{0.27,0.55,0.45}
\definecolor{aquamarine}{rgb}{0.50,1.00,0.83}
\definecolor{azure1}{rgb}{0.94,1.00,1.00}
\definecolor{azure2}{rgb}{0.88,0.93,0.93}
\definecolor{azure3}{rgb}{0.76,0.80,0.80}
\definecolor{azure4}{rgb}{0.51,0.55,0.55}
\definecolor{azure}{rgb}{0.94,1.00,1.00}
\definecolor{beige}{rgb}{0.96,0.96,0.86}
\definecolor{bisque1}{rgb}{1.00,0.89,0.77}
\definecolor{bisque2}{rgb}{0.93,0.84,0.72}
\definecolor{bisque3}{rgb}{0.80,0.72,0.62}
\definecolor{bisque4}{rgb}{0.55,0.49,0.42}
\definecolor{bisque}{rgb}{1.00,0.89,0.77}
\definecolor{black}{rgb}{0.00,0.00,0.00}
\definecolor{blanchedalmond}{rgb}{1.00,0.92,0.80}
\definecolor{blue1}{rgb}{0.00,0.00,1.00}
\definecolor{blue2}{rgb}{0.00,0.00,0.93}
\definecolor{blue3}{rgb}{0.00,0.00,0.80}
\definecolor{blue4}{rgb}{0.00,0.00,0.55}
\definecolor{blueviolet}{rgb}{0.54,0.17,0.89}
\definecolor{blue}{rgb}{0.00,0.00,1.00}
\definecolor{brown1}{rgb}{1.00,0.25,0.25}
\definecolor{brown2}{rgb}{0.93,0.23,0.23}
\definecolor{brown3}{rgb}{0.80,0.20,0.20}
\definecolor{brown4}{rgb}{0.55,0.14,0.14}
\definecolor{brown}{rgb}{0.65,0.16,0.16}
\definecolor{burlywood1}{rgb}{1.00,0.83,0.61}
\definecolor{burlywood2}{rgb}{0.93,0.77,0.57}
\definecolor{burlywood3}{rgb}{0.80,0.67,0.49}
\definecolor{burlywood4}{rgb}{0.55,0.45,0.33}
\definecolor{burlywood}{rgb}{0.87,0.72,0.53}
\definecolor{cadetblue}{rgb}{0.37,0.62,0.63}
\definecolor{chartreuse1}{rgb}{0.50,1.00,0.00}
\definecolor{chartreuse2}{rgb}{0.46,0.93,0.00}
\definecolor{chartreuse3}{rgb}{0.40,0.80,0.00}
\definecolor{chartreuse4}{rgb}{0.27,0.55,0.00}
\definecolor{chartreuse}{rgb}{0.50,1.00,0.00}
\definecolor{chocolate1}{rgb}{1.00,0.50,0.14}
\definecolor{chocolate2}{rgb}{0.93,0.46,0.13}
\definecolor{chocolate3}{rgb}{0.80,0.40,0.11}
\definecolor{chocolate4}{rgb}{0.55,0.27,0.07}
\definecolor{chocolate}{rgb}{0.82,0.41,0.12}
\definecolor{coral1}{rgb}{1.00,0.45,0.34}
\definecolor{coral2}{rgb}{0.93,0.42,0.31}
\definecolor{coral3}{rgb}{0.80,0.36,0.27}
\definecolor{coral4}{rgb}{0.55,0.24,0.18}
\definecolor{coral}{rgb}{1.00,0.50,0.31}
\definecolor{cornflowerblue}{rgb}{0.39,0.58,0.93}
\definecolor{cornsilk1}{rgb}{1.00,0.97,0.86}
\definecolor{cornsilk2}{rgb}{0.93,0.91,0.80}
\definecolor{cornsilk3}{rgb}{0.80,0.78,0.69}
\definecolor{cornsilk4}{rgb}{0.55,0.53,0.47}
\definecolor{cornsilk}{rgb}{1.00,0.97,0.86}
\definecolor{cyan1}{rgb}{0.00,1.00,1.00}
\definecolor{cyan2}{rgb}{0.00,0.93,0.93}
\definecolor{cyan3}{rgb}{0.00,0.80,0.80}
\definecolor{cyan4}{rgb}{0.00,0.55,0.55}
\definecolor{cyan}{rgb}{0.00,1.00,1.00}
\definecolor{darkblue}{rgb}{0.00,0.00,0.55}
\definecolor{darkcyan}{rgb}{0.00,0.55,0.55}
\definecolor{darkgoldenrod}{rgb}{0.72,0.53,0.04}
\definecolor{darkgray}{rgb}{0.66,0.66,0.66}
\definecolor{darkgreen}{rgb}{0.00,0.39,0.00}
\definecolor{darkgrey}{rgb}{0.66,0.66,0.66}
\definecolor{darkkhaki}{rgb}{0.74,0.72,0.42}
\definecolor{darkmagenta}{rgb}{0.55,0.00,0.55}
\definecolor{darkolive}{rgb}{0.33,0.42,0.18}
\definecolor{darkorange}{rgb}{1.00,0.55,0.00}
\definecolor{darkorchid}{rgb}{0.60,0.20,0.80}
\definecolor{darkred}{rgb}{0.55,0.00,0.00}
\definecolor{darksalmon}{rgb}{0.91,0.59,0.48}
\definecolor{darksea}{rgb}{0.56,0.74,0.56}
\definecolor{darkslate}{rgb}{0.18,0.31,0.31}
\definecolor{darkslate}{rgb}{0.18,0.31,0.31}
\definecolor{darkslate}{rgb}{0.28,0.24,0.55}
\definecolor{darkturquoise}{rgb}{0.00,0.81,0.82}
\definecolor{darkviolet}{rgb}{0.58,0.00,0.83}
\definecolor{deeppink}{rgb}{1.00,0.08,0.58}
\definecolor{deepsky}{rgb}{0.00,0.75,1.00}
\definecolor{dimgray}{rgb}{0.41,0.41,0.41}
\definecolor{dimgrey}{rgb}{0.41,0.41,0.41}
\definecolor{dodgerblue}{rgb}{0.12,0.56,1.00}
\definecolor{firebrick1}{rgb}{1.00,0.19,0.19}
\definecolor{firebrick2}{rgb}{0.93,0.17,0.17}
\definecolor{firebrick3}{rgb}{0.80,0.15,0.15}
\definecolor{firebrick4}{rgb}{0.55,0.10,0.10}
\definecolor{firebrick}{rgb}{0.70,0.13,0.13}
\definecolor{floralwhite}{rgb}{1.00,0.98,0.94}
\definecolor{forestgreen}{rgb}{0.13,0.55,0.13}
\definecolor{gainsboro}{rgb}{0.86,0.86,0.86}
\definecolor{ghostwhite}{rgb}{0.97,0.97,1.00}
\definecolor{gold1}{rgb}{1.00,0.84,0.00}
\definecolor{gold2}{rgb}{0.93,0.79,0.00}
\definecolor{gold3}{rgb}{0.80,0.68,0.00}
\definecolor{gold4}{rgb}{0.55,0.46,0.00}
\definecolor{goldenrod1}{rgb}{1.00,0.76,0.15}
\definecolor{goldenrod2}{rgb}{0.93,0.71,0.13}
\definecolor{goldenrod3}{rgb}{0.80,0.61,0.11}
\definecolor{goldenrod4}{rgb}{0.55,0.41,0.08}
\definecolor{goldenrod}{rgb}{0.85,0.65,0.13}
\definecolor{gold}{rgb}{1.00,0.84,0.00}
\definecolor{gray0}{rgb}{0.00,0.00,0.00}
\definecolor{gray100}{rgb}{1.00,1.00,1.00}
\definecolor{gray10}{rgb}{0.10,0.10,0.10}
\definecolor{gray11}{rgb}{0.11,0.11,0.11}
\definecolor{gray12}{rgb}{0.12,0.12,0.12}
\definecolor{gray13}{rgb}{0.13,0.13,0.13}
\definecolor{gray14}{rgb}{0.14,0.14,0.14}
\definecolor{gray15}{rgb}{0.15,0.15,0.15}
\definecolor{gray16}{rgb}{0.16,0.16,0.16}
\definecolor{gray17}{rgb}{0.17,0.17,0.17}
\definecolor{gray18}{rgb}{0.18,0.18,0.18}
\definecolor{gray19}{rgb}{0.19,0.19,0.19}
\definecolor{gray1}{rgb}{0.01,0.01,0.01}
\definecolor{gray20}{rgb}{0.20,0.20,0.20}
\definecolor{gray21}{rgb}{0.21,0.21,0.21}
\definecolor{gray22}{rgb}{0.22,0.22,0.22}
\definecolor{gray23}{rgb}{0.23,0.23,0.23}
\definecolor{gray24}{rgb}{0.24,0.24,0.24}
\definecolor{gray25}{rgb}{0.25,0.25,0.25}
\definecolor{gray26}{rgb}{0.26,0.26,0.26}
\definecolor{gray27}{rgb}{0.27,0.27,0.27}
\definecolor{gray28}{rgb}{0.28,0.28,0.28}
\definecolor{gray29}{rgb}{0.29,0.29,0.29}
\definecolor{gray2}{rgb}{0.02,0.02,0.02}
\definecolor{gray30}{rgb}{0.30,0.30,0.30}
\definecolor{gray31}{rgb}{0.31,0.31,0.31}
\definecolor{gray32}{rgb}{0.32,0.32,0.32}
\definecolor{gray33}{rgb}{0.33,0.33,0.33}
\definecolor{gray34}{rgb}{0.34,0.34,0.34}
\definecolor{gray35}{rgb}{0.35,0.35,0.35}
\definecolor{gray36}{rgb}{0.36,0.36,0.36}
\definecolor{gray37}{rgb}{0.37,0.37,0.37}
\definecolor{gray38}{rgb}{0.38,0.38,0.38}
\definecolor{gray39}{rgb}{0.39,0.39,0.39}
\definecolor{gray3}{rgb}{0.03,0.03,0.03}
\definecolor{gray40}{rgb}{0.40,0.40,0.40}
\definecolor{gray41}{rgb}{0.41,0.41,0.41}
\definecolor{gray42}{rgb}{0.42,0.42,0.42}
\definecolor{gray43}{rgb}{0.43,0.43,0.43}
\definecolor{gray44}{rgb}{0.44,0.44,0.44}
\definecolor{gray45}{rgb}{0.45,0.45,0.45}
\definecolor{gray46}{rgb}{0.46,0.46,0.46}
\definecolor{gray47}{rgb}{0.47,0.47,0.47}
\definecolor{gray48}{rgb}{0.48,0.48,0.48}
\definecolor{gray49}{rgb}{0.49,0.49,0.49}
\definecolor{gray4}{rgb}{0.04,0.04,0.04}
\definecolor{gray50}{rgb}{0.50,0.50,0.50}
\definecolor{gray51}{rgb}{0.51,0.51,0.51}
\definecolor{gray52}{rgb}{0.52,0.52,0.52}
\definecolor{gray53}{rgb}{0.53,0.53,0.53}
\definecolor{gray54}{rgb}{0.54,0.54,0.54}
\definecolor{gray55}{rgb}{0.55,0.55,0.55}
\definecolor{gray56}{rgb}{0.56,0.56,0.56}
\definecolor{gray57}{rgb}{0.57,0.57,0.57}
\definecolor{gray58}{rgb}{0.58,0.58,0.58}
\definecolor{gray59}{rgb}{0.59,0.59,0.59}
\definecolor{gray5}{rgb}{0.05,0.05,0.05}
\definecolor{gray60}{rgb}{0.60,0.60,0.60}
\definecolor{gray61}{rgb}{0.61,0.61,0.61}
\definecolor{gray62}{rgb}{0.62,0.62,0.62}
\definecolor{gray63}{rgb}{0.63,0.63,0.63}
\definecolor{gray64}{rgb}{0.64,0.64,0.64}
\definecolor{gray65}{rgb}{0.65,0.65,0.65}
\definecolor{gray66}{rgb}{0.66,0.66,0.66}
\definecolor{gray67}{rgb}{0.67,0.67,0.67}
\definecolor{gray68}{rgb}{0.68,0.68,0.68}
\definecolor{gray69}{rgb}{0.69,0.69,0.69}
\definecolor{gray6}{rgb}{0.06,0.06,0.06}
\definecolor{gray70}{rgb}{0.70,0.70,0.70}
\definecolor{gray71}{rgb}{0.71,0.71,0.71}
\definecolor{gray72}{rgb}{0.72,0.72,0.72}
\definecolor{gray73}{rgb}{0.73,0.73,0.73}
\definecolor{gray74}{rgb}{0.74,0.74,0.74}
\definecolor{gray75}{rgb}{0.75,0.75,0.75}
\definecolor{gray76}{rgb}{0.76,0.76,0.76}
\definecolor{gray77}{rgb}{0.77,0.77,0.77}
\definecolor{gray78}{rgb}{0.78,0.78,0.78}
\definecolor{gray79}{rgb}{0.79,0.79,0.79}
\definecolor{gray7}{rgb}{0.07,0.07,0.07}
\definecolor{gray80}{rgb}{0.80,0.80,0.80}
\definecolor{gray81}{rgb}{0.81,0.81,0.81}
\definecolor{gray82}{rgb}{0.82,0.82,0.82}
\definecolor{gray83}{rgb}{0.83,0.83,0.83}
\definecolor{gray84}{rgb}{0.84,0.84,0.84}
\definecolor{gray85}{rgb}{0.85,0.85,0.85}
\definecolor{gray86}{rgb}{0.86,0.86,0.86}
\definecolor{gray87}{rgb}{0.87,0.87,0.87}
\definecolor{gray88}{rgb}{0.88,0.88,0.88}
\definecolor{gray89}{rgb}{0.89,0.89,0.89}
\definecolor{gray8}{rgb}{0.08,0.08,0.08}
\definecolor{gray90}{rgb}{0.90,0.90,0.90}
\definecolor{gray91}{rgb}{0.91,0.91,0.91}
\definecolor{gray92}{rgb}{0.92,0.92,0.92}
\definecolor{gray93}{rgb}{0.93,0.93,0.93}
\definecolor{gray94}{rgb}{0.94,0.94,0.94}
\definecolor{gray95}{rgb}{0.95,0.95,0.95}
\definecolor{gray96}{rgb}{0.96,0.96,0.96}
\definecolor{gray97}{rgb}{0.97,0.97,0.97}
\definecolor{gray98}{rgb}{0.98,0.98,0.98}
\definecolor{gray99}{rgb}{0.99,0.99,0.99}
\definecolor{gray9}{rgb}{0.09,0.09,0.09}
\definecolor{gray}{rgb}{0.75,0.75,0.75}
\definecolor{green1}{rgb}{0.00,1.00,0.00}
\definecolor{green2}{rgb}{0.00,0.93,0.00}
\definecolor{green3}{rgb}{0.00,0.80,0.00}
\definecolor{green4}{rgb}{0.00,0.55,0.00}
\definecolor{greenyellow}{rgb}{0.68,1.00,0.18}
\definecolor{green}{rgb}{0.00,1.00,0.00}
\definecolor{grey0}{rgb}{0.00,0.00,0.00}
\definecolor{grey100}{rgb}{1.00,1.00,1.00}
\definecolor{grey10}{rgb}{0.10,0.10,0.10}
\definecolor{grey11}{rgb}{0.11,0.11,0.11}
\definecolor{grey12}{rgb}{0.12,0.12,0.12}
\definecolor{grey13}{rgb}{0.13,0.13,0.13}
\definecolor{grey14}{rgb}{0.14,0.14,0.14}
\definecolor{grey15}{rgb}{0.15,0.15,0.15}
\definecolor{grey16}{rgb}{0.16,0.16,0.16}
\definecolor{grey17}{rgb}{0.17,0.17,0.17}
\definecolor{grey18}{rgb}{0.18,0.18,0.18}
\definecolor{grey19}{rgb}{0.19,0.19,0.19}
\definecolor{grey1}{rgb}{0.01,0.01,0.01}
\definecolor{grey20}{rgb}{0.20,0.20,0.20}
\definecolor{grey21}{rgb}{0.21,0.21,0.21}
\definecolor{grey22}{rgb}{0.22,0.22,0.22}
\definecolor{grey23}{rgb}{0.23,0.23,0.23}
\definecolor{grey24}{rgb}{0.24,0.24,0.24}
\definecolor{grey25}{rgb}{0.25,0.25,0.25}
\definecolor{grey26}{rgb}{0.26,0.26,0.26}
\definecolor{grey27}{rgb}{0.27,0.27,0.27}
\definecolor{grey28}{rgb}{0.28,0.28,0.28}
\definecolor{grey29}{rgb}{0.29,0.29,0.29}
\definecolor{grey2}{rgb}{0.02,0.02,0.02}
\definecolor{grey30}{rgb}{0.30,0.30,0.30}
\definecolor{grey31}{rgb}{0.31,0.31,0.31}
\definecolor{grey32}{rgb}{0.32,0.32,0.32}
\definecolor{grey33}{rgb}{0.33,0.33,0.33}
\definecolor{grey34}{rgb}{0.34,0.34,0.34}
\definecolor{grey35}{rgb}{0.35,0.35,0.35}
\definecolor{grey36}{rgb}{0.36,0.36,0.36}
\definecolor{grey37}{rgb}{0.37,0.37,0.37}
\definecolor{grey38}{rgb}{0.38,0.38,0.38}
\definecolor{grey39}{rgb}{0.39,0.39,0.39}
\definecolor{grey3}{rgb}{0.03,0.03,0.03}
\definecolor{grey40}{rgb}{0.40,0.40,0.40}
\definecolor{grey41}{rgb}{0.41,0.41,0.41}
\definecolor{grey42}{rgb}{0.42,0.42,0.42}
\definecolor{grey43}{rgb}{0.43,0.43,0.43}
\definecolor{grey44}{rgb}{0.44,0.44,0.44}
\definecolor{grey45}{rgb}{0.45,0.45,0.45}
\definecolor{grey46}{rgb}{0.46,0.46,0.46}
\definecolor{grey47}{rgb}{0.47,0.47,0.47}
\definecolor{grey48}{rgb}{0.48,0.48,0.48}
\definecolor{grey49}{rgb}{0.49,0.49,0.49}
\definecolor{grey4}{rgb}{0.04,0.04,0.04}
\definecolor{grey50}{rgb}{0.50,0.50,0.50}
\definecolor{grey51}{rgb}{0.51,0.51,0.51}
\definecolor{grey52}{rgb}{0.52,0.52,0.52}
\definecolor{grey53}{rgb}{0.53,0.53,0.53}
\definecolor{grey54}{rgb}{0.54,0.54,0.54}
\definecolor{grey55}{rgb}{0.55,0.55,0.55}
\definecolor{grey56}{rgb}{0.56,0.56,0.56}
\definecolor{grey57}{rgb}{0.57,0.57,0.57}
\definecolor{grey58}{rgb}{0.58,0.58,0.58}
\definecolor{grey59}{rgb}{0.59,0.59,0.59}
\definecolor{grey5}{rgb}{0.05,0.05,0.05}
\definecolor{grey60}{rgb}{0.60,0.60,0.60}
\definecolor{grey61}{rgb}{0.61,0.61,0.61}
\definecolor{grey62}{rgb}{0.62,0.62,0.62}
\definecolor{grey63}{rgb}{0.63,0.63,0.63}
\definecolor{grey64}{rgb}{0.64,0.64,0.64}
\definecolor{grey65}{rgb}{0.65,0.65,0.65}
\definecolor{grey66}{rgb}{0.66,0.66,0.66}
\definecolor{grey67}{rgb}{0.67,0.67,0.67}
\definecolor{grey68}{rgb}{0.68,0.68,0.68}
\definecolor{grey69}{rgb}{0.69,0.69,0.69}
\definecolor{grey6}{rgb}{0.06,0.06,0.06}
\definecolor{grey70}{rgb}{0.70,0.70,0.70}
\definecolor{grey71}{rgb}{0.71,0.71,0.71}
\definecolor{grey72}{rgb}{0.72,0.72,0.72}
\definecolor{grey73}{rgb}{0.73,0.73,0.73}
\definecolor{grey74}{rgb}{0.74,0.74,0.74}
\definecolor{grey75}{rgb}{0.75,0.75,0.75}
\definecolor{grey76}{rgb}{0.76,0.76,0.76}
\definecolor{grey77}{rgb}{0.77,0.77,0.77}
\definecolor{grey78}{rgb}{0.78,0.78,0.78}
\definecolor{grey79}{rgb}{0.79,0.79,0.79}
\definecolor{grey7}{rgb}{0.07,0.07,0.07}
\definecolor{grey80}{rgb}{0.80,0.80,0.80}
\definecolor{grey81}{rgb}{0.81,0.81,0.81}
\definecolor{grey82}{rgb}{0.82,0.82,0.82}
\definecolor{grey83}{rgb}{0.83,0.83,0.83}
\definecolor{grey84}{rgb}{0.84,0.84,0.84}
\definecolor{grey85}{rgb}{0.85,0.85,0.85}
\definecolor{grey86}{rgb}{0.86,0.86,0.86}
\definecolor{grey87}{rgb}{0.87,0.87,0.87}
\definecolor{grey88}{rgb}{0.88,0.88,0.88}
\definecolor{grey89}{rgb}{0.89,0.89,0.89}
\definecolor{grey8}{rgb}{0.08,0.08,0.08}
\definecolor{grey90}{rgb}{0.90,0.90,0.90}
\definecolor{grey91}{rgb}{0.91,0.91,0.91}
\definecolor{grey92}{rgb}{0.92,0.92,0.92}
\definecolor{grey93}{rgb}{0.93,0.93,0.93}
\definecolor{grey94}{rgb}{0.94,0.94,0.94}
\definecolor{grey95}{rgb}{0.95,0.95,0.95}
\definecolor{grey96}{rgb}{0.96,0.96,0.96}
\definecolor{grey97}{rgb}{0.97,0.97,0.97}
\definecolor{grey98}{rgb}{0.98,0.98,0.98}
\definecolor{grey99}{rgb}{0.99,0.99,0.99}
\definecolor{grey9}{rgb}{0.09,0.09,0.09}
\definecolor{grey}{rgb}{0.75,0.75,0.75}
\definecolor{honeydew1}{rgb}{0.94,1.00,0.94}
\definecolor{honeydew2}{rgb}{0.88,0.93,0.88}
\definecolor{honeydew3}{rgb}{0.76,0.80,0.76}
\definecolor{honeydew4}{rgb}{0.51,0.55,0.51}
\definecolor{honeydew}{rgb}{0.94,1.00,0.94}
\definecolor{hotpink}{rgb}{1.00,0.41,0.71}
\definecolor{indianred}{rgb}{0.80,0.36,0.36}
\definecolor{ivory1}{rgb}{1.00,1.00,0.94}
\definecolor{ivory2}{rgb}{0.93,0.93,0.88}
\definecolor{ivory3}{rgb}{0.80,0.80,0.76}
\definecolor{ivory4}{rgb}{0.55,0.55,0.51}
\definecolor{ivory}{rgb}{1.00,1.00,0.94}
\definecolor{khaki1}{rgb}{1.00,0.96,0.56}
\definecolor{khaki2}{rgb}{0.93,0.90,0.52}
\definecolor{khaki3}{rgb}{0.80,0.78,0.45}
\definecolor{khaki4}{rgb}{0.55,0.53,0.31}
\definecolor{khaki}{rgb}{0.94,0.90,0.55}
\definecolor{lavenderblush}{rgb}{1.00,0.94,0.96}
\definecolor{lavender}{rgb}{0.90,0.90,0.98}
\definecolor{lawngreen}{rgb}{0.49,0.99,0.00}
\definecolor{lemonchiffon}{rgb}{1.00,0.98,0.80}
\definecolor{lightblue}{rgb}{0.68,0.85,0.90}
\definecolor{lightcoral}{rgb}{0.94,0.50,0.50}
\definecolor{lightcyan}{rgb}{0.88,1.00,1.00}
\definecolor{lightgoldenrod}{rgb}{0.93,0.87,0.51}
\definecolor{lightgoldenrod}{rgb}{0.98,0.98,0.82}
\definecolor{lightgray}{rgb}{0.83,0.83,0.83}
\definecolor{lightgreen}{rgb}{0.56,0.93,0.56}
\definecolor{lightgrey}{rgb}{0.83,0.83,0.83}
\definecolor{lightpink}{rgb}{1.00,0.71,0.76}
\definecolor{lightsalmon}{rgb}{1.00,0.63,0.48}
\definecolor{lightsea}{rgb}{0.13,0.70,0.67}
\definecolor{lightsky}{rgb}{0.53,0.81,0.98}
\definecolor{lightslate}{rgb}{0.47,0.53,0.60}
\definecolor{lightslate}{rgb}{0.47,0.53,0.60}
\definecolor{lightslate}{rgb}{0.52,0.44,1.00}
\definecolor{lightsteel}{rgb}{0.69,0.77,0.87}
\definecolor{lightyellow}{rgb}{1.00,1.00,0.88}
\definecolor{limegreen}{rgb}{0.20,0.80,0.20}
\definecolor{linen}{rgb}{0.98,0.94,0.90}
\definecolor{magenta1}{rgb}{1.00,0.00,1.00}
\definecolor{magenta2}{rgb}{0.93,0.00,0.93}
\definecolor{magenta3}{rgb}{0.80,0.00,0.80}
\definecolor{magenta4}{rgb}{0.55,0.00,0.55}
\definecolor{magenta}{rgb}{1.00,0.00,1.00}
\definecolor{maroon1}{rgb}{1.00,0.20,0.70}
\definecolor{maroon2}{rgb}{0.93,0.19,0.65}
\definecolor{maroon3}{rgb}{0.80,0.16,0.56}
\definecolor{maroon4}{rgb}{0.55,0.11,0.38}
\definecolor{maroon}{rgb}{0.69,0.19,0.38}
\definecolor{mediumaquamarine}{rgb}{0.40,0.80,0.67}
\definecolor{mediumblue}{rgb}{0.00,0.00,0.80}
\definecolor{mediumorchid}{rgb}{0.73,0.33,0.83}
\definecolor{mediumpurple}{rgb}{0.58,0.44,0.86}
\definecolor{mediumsea}{rgb}{0.24,0.70,0.44}
\definecolor{mediumslate}{rgb}{0.48,0.41,0.93}
\definecolor{mediumspring}{rgb}{0.00,0.98,0.60}
\definecolor{mediumturquoise}{rgb}{0.28,0.82,0.80}
\definecolor{mediumviolet}{rgb}{0.78,0.08,0.52}
\definecolor{midnightblue}{rgb}{0.10,0.10,0.44}
\definecolor{mintcream}{rgb}{0.96,1.00,0.98}
\definecolor{mistyrose}{rgb}{1.00,0.89,0.88}
\definecolor{moccasin}{rgb}{1.00,0.89,0.71}
\definecolor{navajowhite}{rgb}{1.00,0.87,0.68}
\definecolor{navyblue}{rgb}{0.00,0.00,0.50}
\definecolor{navy}{rgb}{0.00,0.00,0.50}
\definecolor{oldlace}{rgb}{0.99,0.96,0.90}
\definecolor{olivedrab}{rgb}{0.42,0.56,0.14}
\definecolor{orange1}{rgb}{1.00,0.65,0.00}
\definecolor{orange2}{rgb}{0.93,0.60,0.00}
\definecolor{orange3}{rgb}{0.80,0.52,0.00}
\definecolor{orange4}{rgb}{0.55,0.35,0.00}
\definecolor{orangered}{rgb}{1.00,0.27,0.00}
\definecolor{orange}{rgb}{1.00,0.65,0.00}
\definecolor{orchid1}{rgb}{1.00,0.51,0.98}
\definecolor{orchid2}{rgb}{0.93,0.48,0.91}
\definecolor{orchid3}{rgb}{0.80,0.41,0.79}
\definecolor{orchid4}{rgb}{0.55,0.28,0.54}
\definecolor{orchid}{rgb}{0.85,0.44,0.84}
\definecolor{palegoldenrod}{rgb}{0.93,0.91,0.67}
\definecolor{palegreen}{rgb}{0.60,0.98,0.60}
\definecolor{paleturquoise}{rgb}{0.69,0.93,0.93}
\definecolor{paleviolet}{rgb}{0.86,0.44,0.58}
\definecolor{papayawhip}{rgb}{1.00,0.94,0.84}
\definecolor{peachpuff}{rgb}{1.00,0.85,0.73}
\definecolor{peru}{rgb}{0.80,0.52,0.25}
\definecolor{pink1}{rgb}{1.00,0.71,0.77}
\definecolor{pink2}{rgb}{0.93,0.66,0.72}
\definecolor{pink3}{rgb}{0.80,0.57,0.62}
\definecolor{pink4}{rgb}{0.55,0.39,0.42}
\definecolor{pink}{rgb}{1.00,0.75,0.80}
\definecolor{plum1}{rgb}{1.00,0.73,1.00}
\definecolor{plum2}{rgb}{0.93,0.68,0.93}
\definecolor{plum3}{rgb}{0.80,0.59,0.80}
\definecolor{plum4}{rgb}{0.55,0.40,0.55}
\definecolor{plum}{rgb}{0.87,0.63,0.87}
\definecolor{powderblue}{rgb}{0.69,0.88,0.90}
\definecolor{purple1}{rgb}{0.61,0.19,1.00}
\definecolor{purple2}{rgb}{0.57,0.17,0.93}
\definecolor{purple3}{rgb}{0.49,0.15,0.80}
\definecolor{purple4}{rgb}{0.33,0.10,0.55}
\definecolor{purple}{rgb}{0.63,0.13,0.94}
\definecolor{red1}{rgb}{1.00,0.00,0.00}
\definecolor{red2}{rgb}{0.93,0.00,0.00}
\definecolor{red3}{rgb}{0.80,0.00,0.00}
\definecolor{red4}{rgb}{0.55,0.00,0.00}
\definecolor{red}{rgb}{1.00,0.00,0.00}
\definecolor{rosybrown}{rgb}{0.74,0.56,0.56}
\definecolor{royalblue}{rgb}{0.25,0.41,0.88}
\definecolor{saddlebrown}{rgb}{0.55,0.27,0.07}
\definecolor{salmon1}{rgb}{1.00,0.55,0.41}
\definecolor{salmon2}{rgb}{0.93,0.51,0.38}
\definecolor{salmon3}{rgb}{0.80,0.44,0.33}
\definecolor{salmon4}{rgb}{0.55,0.30,0.22}
\definecolor{salmon}{rgb}{0.98,0.50,0.45}
\definecolor{sandybrown}{rgb}{0.96,0.64,0.38}
\definecolor{seagreen}{rgb}{0.18,0.55,0.34}
\definecolor{seashell1}{rgb}{1.00,0.96,0.93}
\definecolor{seashell2}{rgb}{0.93,0.90,0.87}
\definecolor{seashell3}{rgb}{0.80,0.77,0.75}
\definecolor{seashell4}{rgb}{0.55,0.53,0.51}
\definecolor{seashell}{rgb}{1.00,0.96,0.93}
\definecolor{sienna1}{rgb}{1.00,0.51,0.28}
\definecolor{sienna2}{rgb}{0.93,0.47,0.26}
\definecolor{sienna3}{rgb}{0.80,0.41,0.22}
\definecolor{sienna4}{rgb}{0.55,0.28,0.15}
\definecolor{sienna}{rgb}{0.63,0.32,0.18}
\definecolor{skyblue}{rgb}{0.53,0.81,0.92}
\definecolor{slateblue}{rgb}{0.42,0.35,0.80}
\definecolor{slategray}{rgb}{0.44,0.50,0.56}
\definecolor{slategrey}{rgb}{0.44,0.50,0.56}
\definecolor{snow1}{rgb}{1.00,0.98,0.98}
\definecolor{snow2}{rgb}{0.93,0.91,0.91}
\definecolor{snow3}{rgb}{0.80,0.79,0.79}
\definecolor{snow4}{rgb}{0.55,0.54,0.54}
\definecolor{snow}{rgb}{1.00,0.98,0.98}
\definecolor{springgreen}{rgb}{0.00,1.00,0.50}
\definecolor{steelblue}{rgb}{0.27,0.51,0.71}
\definecolor{tan1}{rgb}{1.00,0.65,0.31}
\definecolor{tan2}{rgb}{0.93,0.60,0.29}
\definecolor{tan3}{rgb}{0.80,0.52,0.25}
\definecolor{tan4}{rgb}{0.55,0.35,0.17}
\definecolor{tan}{rgb}{0.82,0.71,0.55}
\definecolor{thistle1}{rgb}{1.00,0.88,1.00}
\definecolor{thistle2}{rgb}{0.93,0.82,0.93}
\definecolor{thistle3}{rgb}{0.80,0.71,0.80}
\definecolor{thistle4}{rgb}{0.55,0.48,0.55}
\definecolor{thistle}{rgb}{0.85,0.75,0.85}
\definecolor{tomato1}{rgb}{1.00,0.39,0.28}
\definecolor{tomato2}{rgb}{0.93,0.36,0.26}
\definecolor{tomato3}{rgb}{0.80,0.31,0.22}
\definecolor{tomato4}{rgb}{0.55,0.21,0.15}
\definecolor{tomato}{rgb}{1.00,0.39,0.28}
\definecolor{turquoise1}{rgb}{0.00,0.96,1.00}
\definecolor{turquoise2}{rgb}{0.00,0.90,0.93}
\definecolor{turquoise3}{rgb}{0.00,0.77,0.80}
\definecolor{turquoise4}{rgb}{0.00,0.53,0.55}
\definecolor{turquoise}{rgb}{0.25,0.88,0.82}
\definecolor{violetred}{rgb}{0.82,0.13,0.56}
\definecolor{violet}{rgb}{0.93,0.51,0.93}
\definecolor{wheat1}{rgb}{1.00,0.91,0.73}
\definecolor{wheat2}{rgb}{0.93,0.85,0.68}
\definecolor{wheat3}{rgb}{0.80,0.73,0.59}
\definecolor{wheat4}{rgb}{0.55,0.49,0.40}
\definecolor{wheat}{rgb}{0.96,0.87,0.70}
\definecolor{whitesmoke}{rgb}{0.96,0.96,0.96}
\definecolor{white}{rgb}{1.00,1.00,1.00}
\definecolor{yellow1}{rgb}{1.00,1.00,0.00}
\definecolor{yellow2}{rgb}{0.93,0.93,0.00}
\definecolor{yellow3}{rgb}{0.80,0.80,0.00}
\definecolor{yellow4}{rgb}{0.55,0.55,0.00}
\definecolor{yellowgreen}{rgb}{0.60,0.80,0.20}
\definecolor{yellow}{rgb}{1.00,1.00,0.00}

\usepackage{graphicx}
\usepackage{times} 
\usepackage{amssymb}
\usepackage{amsmath}
\usepackage{lscape}
\usepackage{url}
\newif\ifAMStwofonts
\AMStwofontstrue

\def\apj{ApJ}
\def\mnras{MNRAS}
\def\nat{Nat}

\def\aap{A\&A}                   % "Astron. Astrophys."
\def\aaps{A\&AS}                 % "Astron. Astrophys. Suppl. Ser."
\def\aj{AJ}                      % "Astron. J."
\def\apjl{ApJ}                   % letter at ApJ

\begin{document}

\title[SS433 before, during and after a major flare]
{SS433's accretion disc, wind and jets: before, during and after a major flare}
\author[Blundell, Schmidtobreick \& Trushkin]
{\parbox[]{6.in} {Katherine M.\ Blundell$^{1}$,  Linda Schmidtobreick$^{2}$ and Sergei Trushkin$^{3}$}\\\\
  \footnotesize
 $^{1}$University of Oxford, Astrophysics, Keble Road, Oxford OX1 3RH\\
 $^{2}$European Southern Observatory, Vitacura, Alonso de Cordova, Santiago, Chile\\
 $^{3}$Special Astrophysical Observatory RAS, Karachaevo-Cherkassian Republic, Nizhnij Arkhyz 36916, Russia}
\maketitle

\begin{abstract} 
  The Galactic microquasar SS433 occasionally exhibits a major flare
  when the intensity of its emission increases significantly and rapidly.  We present an analysis of
  high-resolution, almost-nightly optical spectra obtained before,
  during and after a major flare, whose complex emission lines
    are deconstructed into single gaussians and demonstrate the
    different modes of mass loss in the SS433 system.  During our
  monitoring, an initial period of quiescence was followed by
  increased activity which culminated in a radio flare. In the
  transition period the accretion disc of SS433 became visible in
    H$\alpha$ and He\,I emission lines and remained so until the
  observations were terminated; the line-of-sight velocity of the
  centre of the disc lines during this time behaved as though
  the binary orbit has significant eccentricity rather than
  being circular, consistent with three recent lines of
  evidence. After the accretion disc appeared its rotation speed, as
  measured by the separation of the H$\alpha$ disc emission lines,
  increased steadily from 500\,km/s to 700\,km/s. The launch speed of
  the jets first decreased then suddenly increased. At the same time
  as the jet launch speed increased, the wind from the accretion disc
  doubled in speed.  Two days afterwards, the radio flux
  exhibited a flare.  These data suggest that a massive ejection of
  material from the companion star loaded the accretion disc and the
  system responded with mass loss via different modes that
  together comprise the flare phenomena.  We find that archival
  data reveal similar behaviour, in that when the measured jet launch
  speed exceeds 0.29\,$c$ this is invariably simultaneous with, or a
  few days before, a radio flare.  Thus we surmise that a major flare consists of the overloading of the accretion disc, resulting in the speeding up of the H-alpha rotation disc lines, followed by enhanced mass loss not just via its famous jets at higher-than-usual speeds but also directly from its accretion disc's wind.
  \end{abstract}


\section{Introduction}

The Galactic microquasar SS433 is famous for its continual ejection of
plasma collimated in two oppositely-directed jets launched at speeds
that average over time to one quarter of the speed of light.  The
system is a binary with a period of 13.08 days
\citep{Cramptonetal1980} that eclipses at both oppositions
\citep[e.g.][]{Goranskiietal98}. There is evidence that He II 4486
\AA\ emission has been observed from the base of the jets
\citep{Crampton1981,Fabrika1990} and C II lines orbiting with the
compact object have been detected (Gies et al 2002).  The accretion
disc is not always revealed by optical observations \citep[although
see][]{falamo1987,Perez2010} but the signature of the accretion disc,
namely a widely-separated pair of emission lines, appears to be
prominent in the near infra-red \citep{Perez2009}.  The system is
rather massive with the compact object and disc having a mass of
$\sim$16\,M$_\odot$ and the companion star having $\sim$24\,M$_\odot$.
See \cite{BBS2008}; \cite{Fabrika2004} for a historic review.

In August 2004 a campaign of nightly observations of SS433 was
initiated with the NTT 3.5-m telescope on La Silla,
Chile\footnote{Based on observations collected at the European
  Organisation for Astronomical Research in the Southern Hemisphere,
  Chile (Program ID: 273.C-5050)}. The observations continued until
early November.  There were few gaps in the coverage and this sequence
of nightly observations reveal very rich behaviour in this
microquasar. The results obtained on the relativistic jets
\citep{BBS2007} and on the properties of the stationary emission lines
during the first half of the series of observations \citep{BBS2008}
have already been published. This paper is concerned with the
observations during the second half of the series, when SS433 was
switching from a quiescent state to a period of activity culminating
in a major flare.

\subsection{On the physical interpretation of the two types of flare}

A number of authors \citep[for example, ][]{Seaquist1982,Fabrika2004}
have suggested that SS433 exhibits two types of radio flare: (1) those
where the flux density at all frequencies rises and maintains a
constant, although fairly flat spectrum ($S_{\nu} \propto \nu^{-0.2}$)
and (2) those where a peak is seen at frequencies of several GHz which
then gradually moves to lower frequencies over a timescale of a few
days.  Time-resolved radio studies \citep{Feidler1987,Bonsig1986} have
revealed clusters of flares separated by periods of quiescent
emission, without any indication of significant stable periodicity
(see discussion in Sec\,\ref{sec:clustering}).  

In observations of the Galactic microquasar Cygnus X-3, 
a major flaring event in September 2001 was resolved into two distinct
peaks in intensity, separated by a couple of days.  The first of these
{\it (a type-1 flare)} has a fairly flat spectrum and, by
milli-arcsecond resolution observations with the Very Long Baseline
Array (VLBA), is seen to correspond to emission from the nucleus
(i.e.\ core) of this microquasar.  The second peak {\it (a type-2
  flare)} is likely to correspond to the appearance of jet ejecta,
specifically their brightening as the bolides expand and transition
from being optically thick to optically thin and subsequent fading as
expansion of the bolides continues \citep{jcamj2004}.  Similarly,
\citet{Vermeulen1993radiospectra}, in their combined radio and optical
study of SS433 in 1987, found two types of flare: one with a peak flux
that is similar over the entire observed range of radio wavelengths
coinciding with a brightening of the core component of the system
\citep{Vermeulen1993vlbi} and the second type of flare where a 
  lower peak flux density at high frequency evolves into a higher peak
  flux at low frequency and in which the same campaign of Very
  Long Baseline Interferometry (VLBI) imaging revealed that the
flaring takes place some distance away from the nucleus.

Separate flaring of the radio emission from the nucleus (perhaps
arising from a disc wind \citep[e.g.\,][]{KZ2007}) and from intense
jet ejecta we believe correspond to there being two distinct modes of
energy and angular momentum output at work, as explored by
\citet{Nipoti2005}.  We remark that \citet{BBS2008} found a
  persistent fast wind to be rooted in SS433's accretion disc and that
  \citet{Perez2009} found from infra-red spectroscopy that this wind
  is the dominant mode of mass-loss in this object.  

We emphasize that not all radio emission in microquasars should be
interpreted as arising from jets, and that radio emission due to disc
winds should be considered especially when there are different
spectral characteristics that could be clues to alternative origins
\citep[c.f.\ ][]{Fender2004}.  A model for nuclear emission in
  radio-quiet quasars arising from optically-thin bremsstrahlung from
  (scale-free) accretion disc winds has been presented by
  \citet{KZ2007}.

\section{Data}

Our nightly spectra were taken from Julian Date 2453000 +245.5 to
+321.5, after which SS433 was not observable from La Silla.  Up to
Day\,+274.5 only one observation was missed and during this period
SS433 was quiescent \citep{BBS2007,BBS2008}. Between Days\,+274.5 and
+287.5 there were only two observations (+281.5, +282.5) and there was
a gap of two nights after +291.5. After Day\,+310.5 there is only one
observation, Day\,+321.5. All spectra were taken with the ESO 3.6-m
New Technology Telescope with the EMMI instrument \citep{dekker1986},
using grating 6 and a 0.5-arcsec slit. The resolution was 2.2\,\AA\ at
6000\,\AA\ and the wavelength range was 5800 to 8700\,\AA.

\begin{figure}
\begin{center}
   \includegraphics[width=8.0cm]{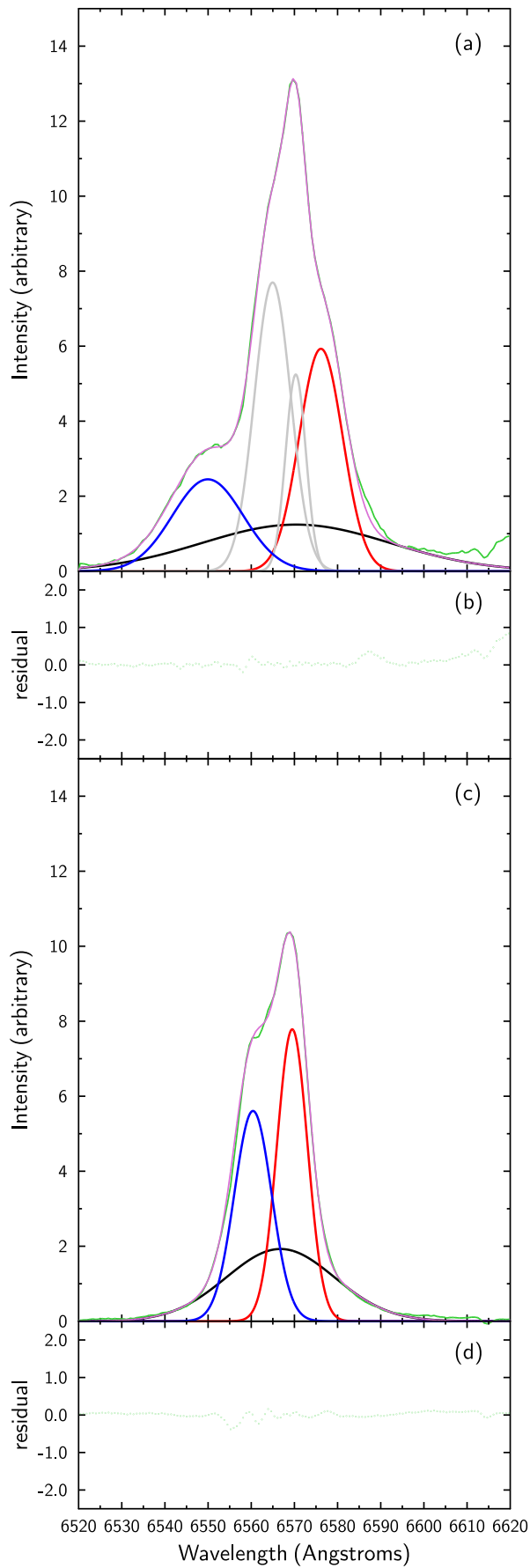} 
   \caption{Examples of our observed spectra of the ``stationary'' H$\alpha$
     feature shown in green from (a) Day +295 and (c) Day +265 with the
     five and three gaussians with which these emission features were
     respectively fitted.  The superpositions of the gaussians fitted
     for each spectrum are plotted in purple;  the difference between
     the green and purple curves --- the residuals to the fits --- are
   shown in panels (b) and (d) respectively.   }
\label{fig:residuals}
\end{center}
\end{figure}

IRAF was used for the basic data reduction including overscan
subtraction, flat-fielding, and wavelength calibration.  No flux
calibration was performed. Instead, the spectra have been normalised
for the continuum using the program SPLOT written and kindly
provided by Jochen Liske. The ``moving'' H$\alpha$ jet lines were
individually fitted with gaussians for all spectra and subtracted to
obtain a clean data set of ``stationary'' H$\alpha$ lines.    
Fig\,\ref{fig:residuals} indicates the high signal-to-noise of the
  gaussians we fitted to the stationary H$\alpha$ complexes.

We do not refer in this paper to the flux of the spectra because
  we do not have absolute flux calibration for these data (although
  the instrumental setup and exposure time were identical for each
  observation, and the weather conditions were fairly similar).  We
  instead use the term intensity to mean the ``area'' of an emission
  line in the spectrum, which is calculated as the height of the
  fitted gaussian multiplied by its FWHM; for our normalised spectra
  this value is the same as the equivalent width of the line. 

\begin{figure}
\begin{center}
   \includegraphics[width=7.6cm]{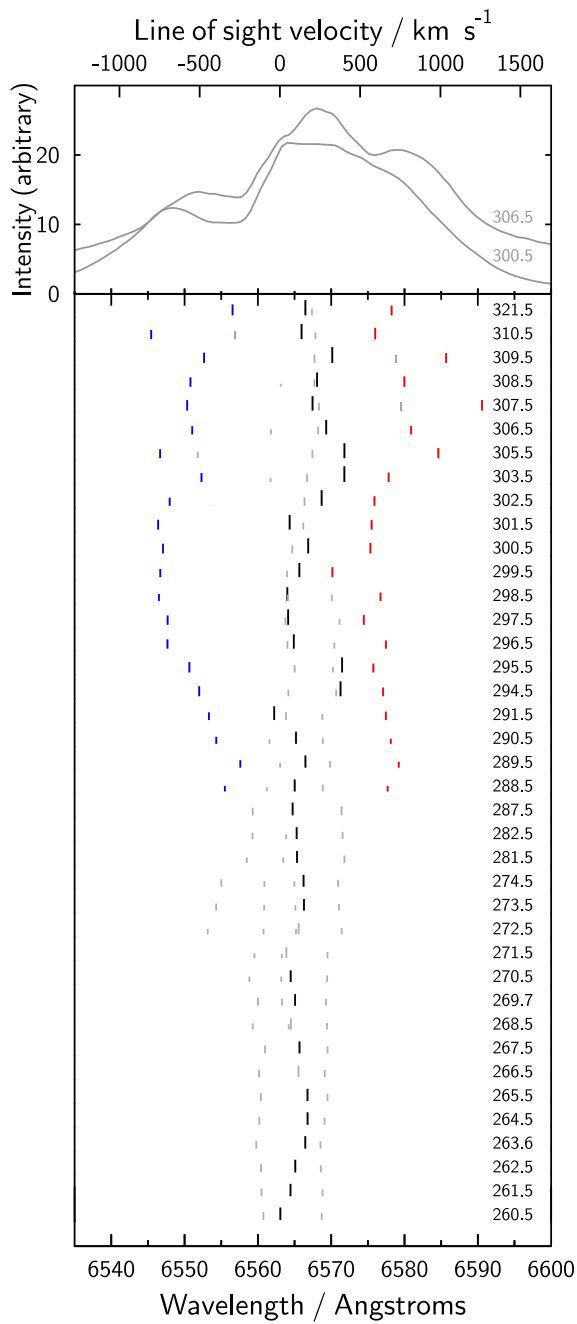} 
   \caption{Wavelengths of the centroids of the gaussian-fitted
     components of the sequence of Balmer H$\alpha$ spectra. Julian
     date increases vertically.  The heights of the lines reflect the
     widths of the gaussian fits, on a logarithmic scale.  The
     signature of the circumbinary disc is clear in the grey lines
     before Day\,+287; a fuller picture of this is in \citet{BBS2008}.
     After Day\,+287 the accretion disc is revealed; the lines
     attributed to the red and blue shifted regions are appropriately
     colour coded. The black lines are broad and associated with the
     wind from the disc; the mean redshift of these lines jumps about
     150\,km\,s$^{-1}$ to the red between Days +291 and +294 and then
     almost back again, and the same behaviour is seen on Days +301
     and +303.  The colours are fitted algorithmically, to avoid
     subjective inference, such that the broad (i.e.\ wind) line in
     each spectrum is coloured black, then the outermost lines are
     coloured red and blue and all remaining lines are grey.  The top
     panel demonstrates clear changes between example spectra from
     Days\,+300 and +306. }
\label{fig:ideagram}
\end{center}
\end{figure}

\begin{figure}
   \centering
   \includegraphics[width=8.5cm]{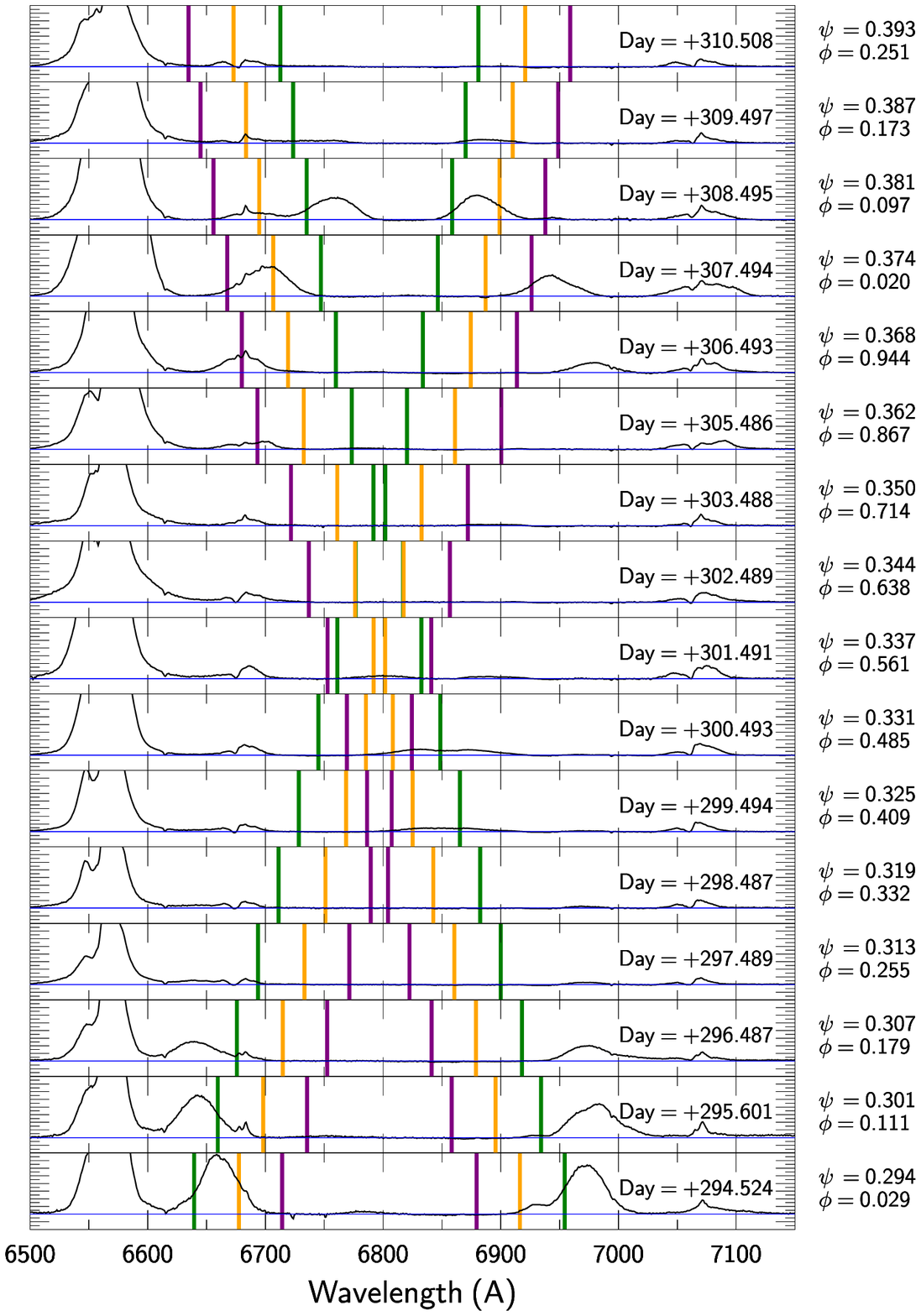} 
   \caption{This figure shows portions of successive optical spectra
     we obtained that illustrate the enhanced nature of the jets'
     emission lines at Day +294, followed by their seeming
     disappearance a few days later, and their subsequent reappearance
     coinciding with orbital phase $\phi \sim$1, almost one full
     orbital period after the intensified jet emission commenced.  The
     coloured vertical lines show the predicted positions of the jet
     lines if they obeyed the standard kinematic model assuming the
     nodding parameters found by Vermeulen (1989)
     with the angle of the precession cone axis to our line-of-sight
     being 78\,deg, with the cone semi-angle being 18\,deg (green),
     19\,deg (orange) and 20\,deg (purple).  The precession phase
     $\psi$ is also indicated, following convention A ({\tt
       www-astro.physics.ox.ac.uk/$\sim$kmb/ss433/}).}
   \label{fig:linesdisappear}
\end{figure}

In Fig.\,\ref{fig:ideagram} we show the wavelengths of the centres of
the components fitted to the stationary H$\alpha$ complex. The data
displayed start at Day\,+260.5 and until Day\,+287.5 most spectra
  are fitted with a single broad line attributed to the wind \citep[on
  the basis of our analysis in][]{BBS2008} (thus, lines with FWHM
  greater than 10\,\AA\ are shown in black) and a pair of lines
showing negligible change in redshift as a function of time 
  (shown in grey) inferred to be the inner rim of the circumbinary
disc (Blundell, Bowler \& Schmidtobreick 2008, see fig.\,1).  A
  striking change occurs at or around Day\,+288.5 when the H$\alpha$
complex is broadened by the addition of components at comparatively
extreme excursions in velocity.  These spectra usually comprised
a single broad line (FWHM 10 to 20 \AA) and two or four lines which
are narrow (FWHMs of a few \AA\ only).  The same behaviour is
seen for the He\,I lines at 6678\,\AA\ and 7065\,\AA\
\citep{linda2006}.  After Day\,+296 the He~I lines and the hydrogen
Paschen series develop pronounced P\,Cygni features and an O~I line at
7772\,\AA\ exhibits absorption to the blue cutting deep into the
continuum.  The O~I line at 8446\,\AA\ does not show any such
features.  The behaviour of these lines will be described in a
  forthcoming paper.

\begin{figure}
\begin{center}
   \includegraphics[width=8.5cm]{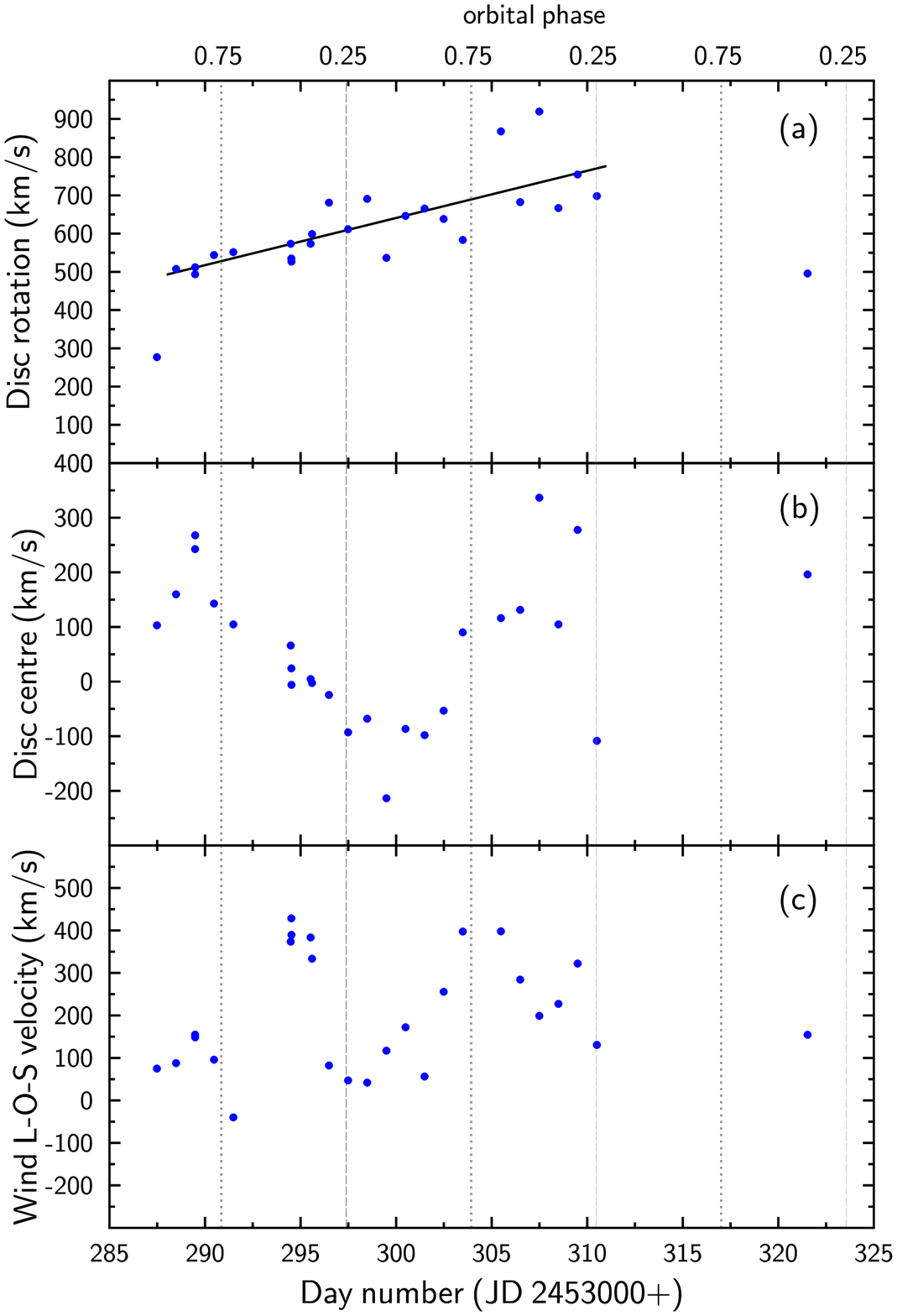} 
   \caption{ 
      {\it (Panel a:)} The velocity with which the
       widely-spaced lines (interpreted as accretion disc lines) orbit
       the compact object, constructed by taking half their
       difference in redshift.  The accretion disc became clearly visible about Day\,+288, 
       although there may be intermittent glimpses earlier. The
     speed with which the H$\alpha$ emitting region orbits the compact
     object appears to increase from about 500 to 700\,km\,s$^{-1}$
     over about 20 days.  
    Also shown is the best-fit straight line for all
    points between Days\,+288 and +311; this line has a gradient of
    $12 \pm 3$\,km\,s$^{-1}$day$^{-1}$ while the gradient of the line
    fitted if the two extreme speed points are excluded is $9 \pm 2$\,km\,s$^{-1}$day$^{-1}$.
    {\it (Panel b:)} The mean line-of-sight velocity of the two extreme
     components of H$\alpha$.  Although the data are barely adequate
     for the purpose, they may echo a 13-day oscillation;  the form of
     this pattern is considered further in Fig\,\ref{fig:eccentric}.  
  {\it (Panel c:)} The line-of-sight velocity of the centre of the
  wind. After Day\,+291 the wind speed doubles (see
  Fig.\,\ref{fig:flaresequence}b) and the mean recessional speed of
  the wind has shifted to the red.  The variation of
    line-of-sight velocity is suggestive of an oscillation whose
    amplitude is consistent with previously reported values
    \citep[e.g.\ ][]{Crampton1981,Fabrika1990}.  For all
  of the above panels, the orbital phases 0.25 and 0.75 are indicated
  as dashed and dotted lines respectively. }
\label{fig:disc}
\end{center}
\end{figure}

\section{The appearance of SS433's accretion disc in the optical}
\label{sec:optical}
The data shown in Fig.\,\ref{fig:ideagram} on the extreme blue and red
features of SS433's stationary H$\alpha$ complex are simply explained
as being radiated from a ring or disc orbiting the compact object
itself, with a speed in excess of 500\,km s$^{-1}$, consistent with
the picture seen in infra-red spectroscopy \citep{Perez2009}; 
  \citet{Bowler2010} has published a subset of our data showing the
  behaviour presented here.  If the H$\alpha$-emitting region of the
accretion disc is perpendicular to the jet axis, then the disc is
close to edge-on to Earth during these observations.   
  Fig.\,\ref{fig:linesdisappear} shows the timing of ``cross-over'' of
  the jet lines: the epoch when the jets are launched in the plane of
  the sky.  This figure shows that variations in {\it the separation of the
  red and blue jet lines} differs from the prediction of the kinematic
  model.  This is a manifestation of the phenomena analysed and
  published in \citet{BB2005} and in \citet{BBS2007}: angular
  variations in the cone angle of rms $\sim3$ degrees have been
  exhibited by this system since its discovery.  For example,
  increasing separation of the jet lines \citep[][eqn\,3]{BB2005}
  indicates a typical instance of the cone angle increasing.

The rotational velocity of the radiating material in the accretion
disc is found by taking by half the difference of the redshifts of the
two components while the line-of-sight velocity of the centre of the
disc is given by the mean of their redshifts, shown in
  Fig\,\ref{fig:disc}a and Fig\,\ref{fig:disc}b respectively as a
  function of time.  The line-of-sight velocities of the disc
material emitting H$\alpha$ are modulated by the orbital velocity of
the compact object about the binary centre of mass, which has been
reported to be $\sim$175\,km\,s$^{-1}$ to $\sim$200\,km\,s$^{-1}$
\citep{Crampton1981,Fabrika1990}.  Given this explanation it is
entirely possible that the extreme blue features observed on
Days\,+273 and 274 are earlier glimpses of the blue rim of the
accretion disc.  The rotational velocity as derived from the
  separation of the H$\alpha$ disc lines (Fig.\,\ref{fig:disc}a) is
about 500 km s$^{-1}$ when this structure is first revealed and
reaches as much as 700\,km s$^{-1}$ at the end of the sequence of
observations.  Given the precession phase (see
  Fig\,\ref{fig:linesdisappear}), the disc is believed to be close to
edge-on at the epoch of these observations and so no significant
fraction of this rotation-speed change can be attributed to
changes in the disc orientation.

The centroid of the pair of widely-separated lines which are so marked
in both H$\alpha$ (Fig.\,\ref{fig:ideagram}) and in He I 7065\,\AA\
reach a (local) maximum in wavelength at Day\,+289.5, a couple of days
before orbital phase 0.75.  The pair reach their most extreme blue
position more than half an orbital period later (but then remain at
approximately the same wavelength for a few days).  After Day\,+301.5
the pair of widely-separated lines returns towards the red.  These two
lines, swinging together in phase, are separated by over
1000\,km\,s$^{-1}$.  During the period after Day\,+287 the jets were
close to the plane of the sky, making an angle to the line of sight of
$80^{\circ}$ initially, then $85^{\circ}$ by Day\,+294; see
\citet{BBS2007} and Fig.\,\ref{fig:linesdisappear}.

In a multiplet of lines emerging from a windy and smoggy system there
is bound to be some uncertainty about the proper assignment of lines
to the opposite sides of the disc, but we have strictly taken the
extreme blue and extreme red components of the H$\alpha$ complex after
Day\,+287.5 and paired them.  It is clear from both
Fig.\,\ref{fig:ideagram} and Fig.\,\ref{fig:disc}b that the simple
explanation advanced above, while representing very well the data over
the first half of the orbital cycle, has inadequacies after
Day\,+298. The lines attributed to the edges of the disc do not return
toward the red along a sinusoidal curve but rather hang in the blue
for several days before snapping back rather abruptly around
Day\,+303.5. It is also clear that the pattern becomes rather noisy
after Day\,+300. These irregularities cannot be attributed to
measurement error (as is evident from Fig\,\ref{fig:residuals})
and are unlikely to be due to inadequate modelling of the line
profiles.  We explore possible explanations for the behaviour
  after Day\,+298 in Section\,\ref{sec:lurch}.

\subsection{A further signature of SS433's binary orbit's eccentricity}
\label{sec:eccentricity}

\begin{figure}
   \centering
   \includegraphics[width=8.5cm]{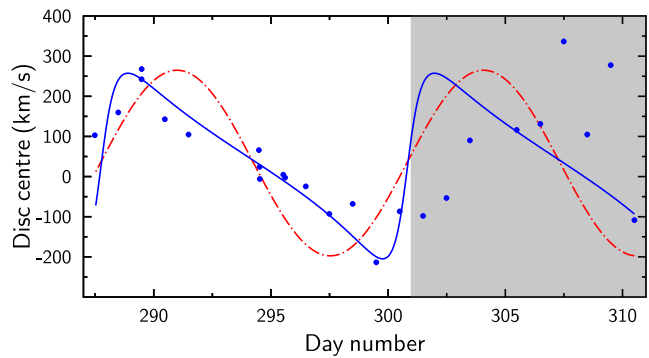} 
   \caption{This figure shows the line-of-sight velocities observed
     when the accretion disc was revealed in the build-up to the flare
     on a white background, and after the flare when the jet
     speed ceases to be high on a grey background.  The orbital
     phases of the maximum and minimum line-of-sight velocities are a
     function of the eccentricity of the orbit and the red dot-dash
     line indicates the predicted line-of-sight velocity if the
     eccentricity were zero (i.e.\ circular orbit) while the blue
     solid line indicates the predicted line-of-sight velocity if the
     eccentricity is 0.55.  During the flare when there seems to be a
     good view of the disc the latter is a better description than the
     former.  Both lines are plotted for the semi-major axis of the
     orbit being oriented along our line-of-sight.  }
   \label{fig:eccentric}
\end{figure}
 
If the orbit of the black hole and accretion disc about the companion
star were circular then the line-of-sight speed of the disc would be
greatest at orbital phase 0.75 (when the compact object is on a
receding path) and least at orbital phase 0.25 (when the compact
object is on an approaching trajectory).  Inspection of
Fig.\,\ref{fig:disc}b shows that the greatest line-of-sight speed of
the disc after it first appears is observed at closer to orbital phase
0.65 while the least line-of-sight speed following this is less clear,
but it is plausibly consistent with orbital phase 0.45.  If correctly
identified, then this is a clear signature that the black hole and
disc are in an elliptical, not a circular, orbit around the companion
star with periastron being at orbital phase 0.5.
Fig.\,\ref{fig:eccentric} shows the predicted line-of-sight velocities
for this system if the eccentricity $e = 0.55$.  Photometric
monitoring of this binary system in no way precludes the orbit being
significantly eccentric but the occurrence of the eclipses at orbital
phases 0 and 0.5 (i.e.\ equally spaced) requires the
semi-major axis to be aligned close to the line of sight.  There are
three other independent indications that the binary orbit of SS433 is
eccentric: (i) the peak jet speed of SS433 varies sinusoidally with
orbital period \citep{BB2005} suggesting that something about the
orbit breaks circular symmetry; (ii) \citet{Perez2009} find that the
constraints on the size of the companion star require the semi-minor
axis to be smaller than the semi-major axis; (iii) the precession of
the radio ruff of SS433 having a periodicity of as short as $\sim 42$
orbital periods requires significant eccentricity in the binary orbit
\citep{Doolin2009}.

Fig.\,\ref{fig:disc}c displays the variation of the line-of-sight
velocity of the centre of the broad Gaussian representing the wind in
the system, established as coming off the accretion disc prior to
Day\,+274.5 \citep{BBS2008}. This centre oscillates in line-of-sight
velocity in a similar way to the lines from the periphery of the disc,
but suffered a displacement in the mean towards the red between
Day\,+291 and +294 and again between Day\,+303 and +305.  This
displacement amounts to approximately 150\,km\,s$^{-1}$; less than one
tenth of the width of the wind lines after Day\,+294.

Thus although Fig.\,\ref{fig:ideagram} and Fig.\,\ref{fig:disc}
display data that are somewhat erratic, these data are sufficient to
establish that the high speed components of the stationary Balmer
H$\alpha$ lines which appear after Day\,+287.5 are contributed by a
ring, or tight spiral, within the accretion disc of SS433.   The
  enhanced wind which appears at the onset of the flare is rooted in
  the accretion disc, just as in more placid times
  \citep[][Sec.\,2.1]{BBS2008}.

\begin{figure*}
\begin{center}
   \includegraphics[width=\textwidth]{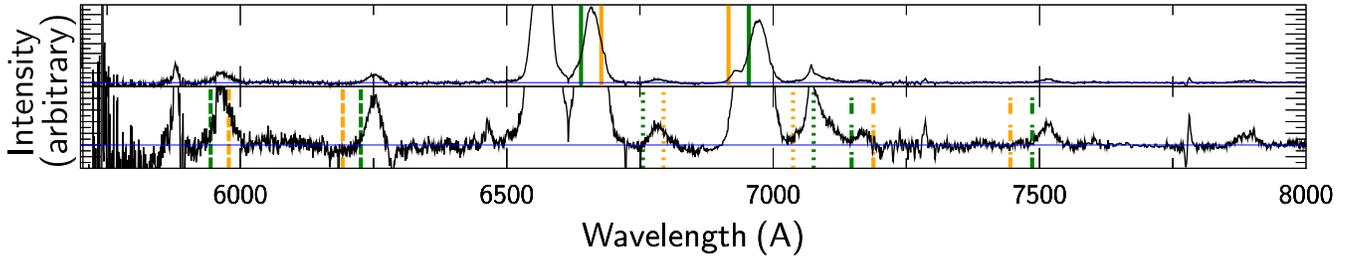} 
   \caption{The spectrum from Day +294 is shown with two
     different  intensity scales to illustrate the moving jet lines in
     H$\alpha$ and in certain helium lines.  The upper panel has solid
     green and gold lines (same parameters as in
     Fig.\,\ref{fig:linesdisappear}) redshifted and blueshifted
     appropriate for rest-frame H$\alpha$; the lower panel shows the
     same coloured lines as dashed for He 5876, dotted for He 6678 and
     dot-dashed for He 7065.  Note that the more redshifted line
     in each pair of ``moving lines'', whether helium or hydrogen, is
     redshifted further than the simple kinematic model predicts; this
     increase in mean redshift is a manifestation of the increased jet
     speed at this time.  }
\label{fig:helium}
\end{center}
\end{figure*}

\subsection{The disruption of the circumbinary disc}

In the days that follow the flare, after accounting for a broad
component of the H$\alpha$ complex as wind and the high velocity
components as the manifestation of the accretion disc, in most cases
there are one or two lines unaccounted for.  It seems likely that some
of these are the continuation of the signature of the circumbinary
disc (albeit at different wavelengths from what is observed prior to
Day\,+294) having possibly undergone some disruption during the flare
outburst.  We assume that given the inherent stability of precessing
circumbinary orbits \citep[][and 2011]{Doolin2009}, this structure
will regroup and form a more cogent signature as it has presumably
done following major flares in the past.

\section{The sequence of events in this flare history}
\label{sec:sequence}

\begin{figure*}
\begin{center}
\includegraphics[width=\textwidth]{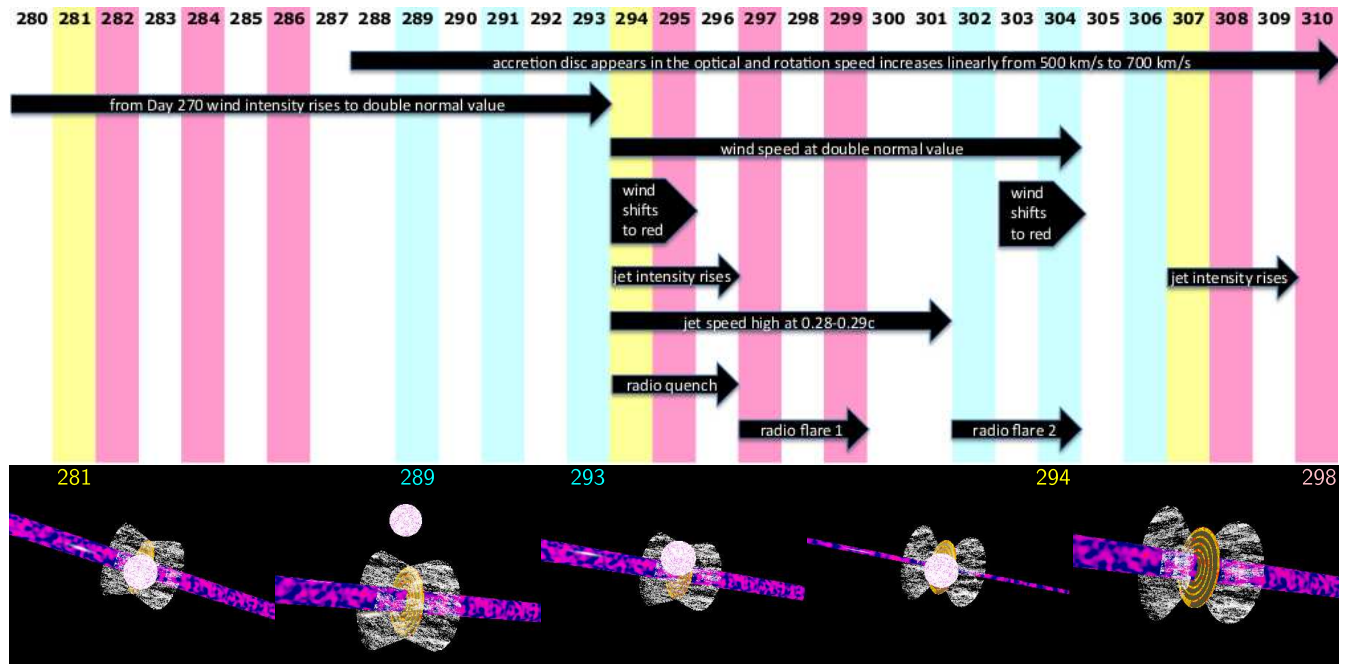}
\caption{{\it Upper panel:} schematic illustration of the various
  phenomena involved in the flaring episode.  Yellow columns indicate
  where the companion star eclipses the compact object and inner part
  of its accretion disc; blue and white alternating columns indicate
  when the black hole and disc are moving towards Earth; pink and
  white alternating columns indicate when the black hole and disc are
  moving away from Earth.  The numbers at the top of the columns are
  Julian Day numbers minus 2453000.  {\it Lower panel:} illustrations
  of the relative locations of the accretion disc, the accretion disc
  wind, the companion star and the jet radius throughout the period of
  observation, for the orbital and precessional phases appropriate to
  the dates illustrated.  The illustrated diameters of the jets act as
  a simple visual aid to remind the reader that the faster jet speed
  observed on Day\,+294 likely comes from a smaller launch radius,
  consistent with the increasing rotation speed of the accretion disc
  observed during these obervations.}
\label{fig:timesequence}
\end{center}
\end{figure*}

\begin{figure}
\begin{center}
   \includegraphics[width=8.5cm]{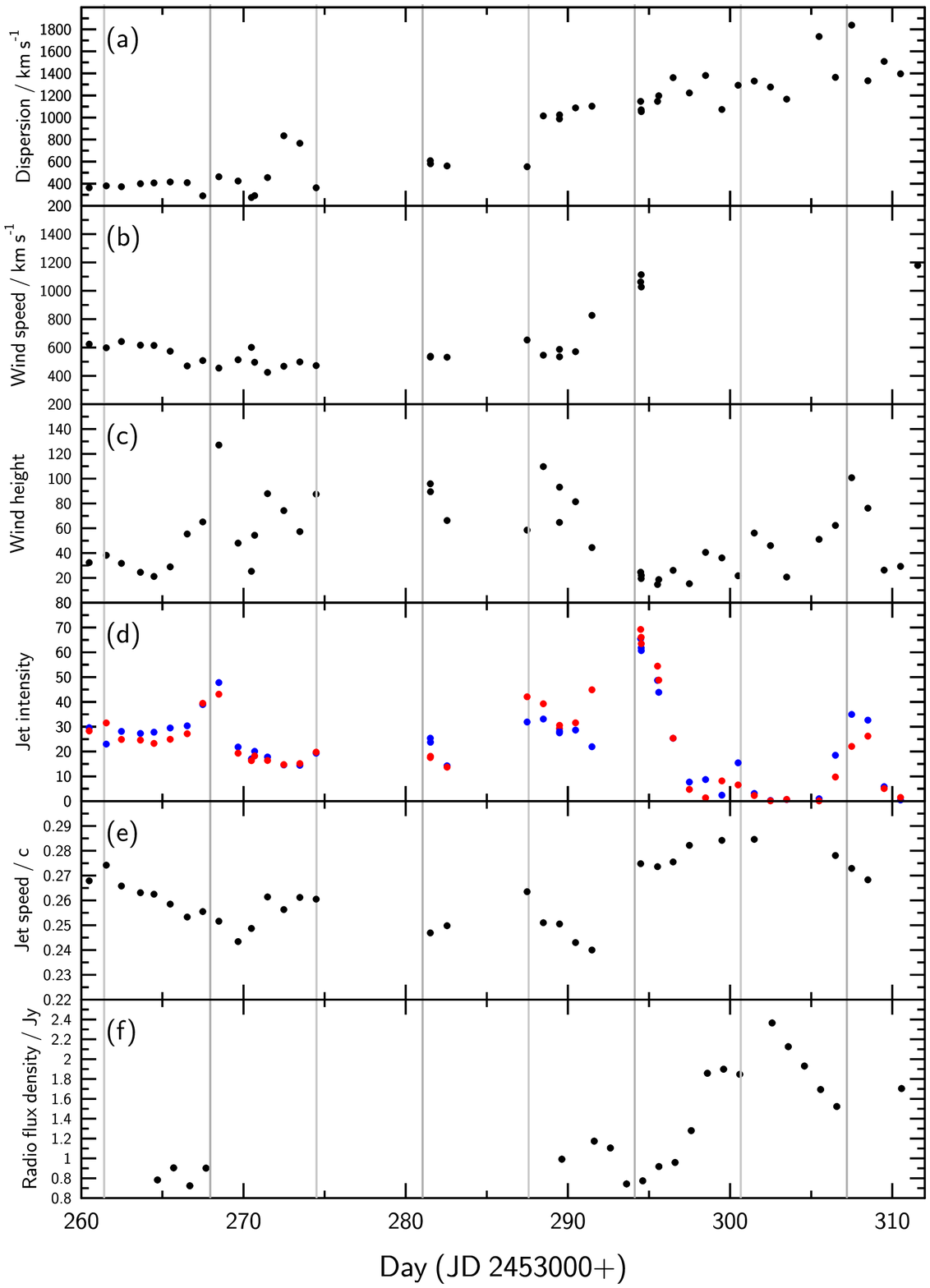} 
   \caption{The time sequence of various phenomena involved in the
     flaring episode. (a) The separation between the extreme red and
     blue components; the disc becomes visible about Day\,+288.  (b)
     The wind speed, as measured by the FWHM of the broadest
     Gaussian. This projected speed is close to the plane of the disc
     and it increases rapidly around Day\,+291 from approximately
     500\,km\,s$^{-1}$ to $> 1000$\,km s$^{-1}$.  (c) The height of
     the wind lines, i.e.\ the height of the broadest component of the
     stationary H$\alpha$ complex.  There appears to be a significant
     drop in the intensity of H$\alpha$ radiation from the wind after
     the jet speed picked up and before the radio flare developed,
     i.e.\ around Day\,+294.  (d) The intensity of the jet lines,
     after the line-free continuum has been fitted and subtracted off.
     (e) The speed of the jets, which are abnormally high from
     Day\,+294 onwards. The jets disappear temporarily between
     Days\,+302 to 305.  (f) The radio flux density at 2.15\,GHz from
     RATAN. The radio flare peaks at Day\,+298 and again at
     Day\,+302.5.  The faint grey lines indicate orbital phases 0 and
     0.5, with day 294 having orbital phase 0. }
\label{fig:flaresequence}
\end{center}
\end{figure}

Fig.\,\ref{fig:timesequence} is a schematic illustration of the events
that collectively comprise the flare phenomenon.
Fig.\,\ref{fig:linesdisappear} shows that the moving H$\alpha$ lines
from the jet ejecta appear strongly on Day\,+294; these continue to be
ejected up to Day\,+297, then largely disappear but appear again on
Day\,+307.  This figure shows that the disappearance is total between
Days\,+302 and 305.  The optical jet lines reappear, rather more
strongly, on Day\,+307 and +308, before disappearing again. It is also
the case that the jet line intensities are falling between Day\,287 up
to Day\,+291 (Fig\,\ref{fig:flaresequence}d) but on Day\,+294 achieve
their greatest values since observations commenced on Day\,+245.  In
fact, the moving He\,I lines reflect this behaviour too: we observe
the same shifts from helium lines with rest wavelengths 5876, 6678 and
7065 (Fig.\,\ref{fig:helium}).  These helium lines appear on Day +294
as shown in this figure but are somewhat faded on Day +295; they
appear with the same shifts as H$\alpha$ on Day +307 and again on Day
+308 when they appear somewhat stronger.
(Fig.\,\ref{fig:flaresequence}d shows that there is a small peak in
jet line intensity around Day\,+268 but this is due to primary eclipse
occurring at this time; this is the extent to which it is likely that
the jet line intensity enhancement we observe on Day\,+294 is due to
primary eclipse.)  The equivalent widths on Day\,+294 are
significantly greater than those observed during primary eclipses
before Day\,+287.  This episode is similar to that observed by
\citet{kopylov1985}, where the jet equivalent widths flared from
invisibility and then collapsed again.  The episode of jet flaring
they observed coincided with a photometrically-determined optical
flare and also coincided with a primary eclipse; we return to the
phasing of flares in Sec\,\ref{sec:clustering}.

In Fig.\,\ref{fig:flaresequence} we display the variation with time of
quantities likely to cast light on the nature of this flaring
episode. The separation of the extreme components of H$\alpha$
suddenly increases at Day\,+288 when the accretion disc is unveiled
(see Fig.\,\ref{fig:flaresequence}a). The speed of the wind from the
disc starts rising about 3 days later and has doubled by Day\,+294
(Fig.\,\ref{fig:flaresequence}b) five days after the disc was
revealed.  The speed of the jets \citep{BBS2007} drops at about the
time the disc was revealed but by Day\,+294 has risen to a rather high
value, $\beta$ $\sim$ 0.28, as shown in
Fig.\,\ref{fig:flaresequence}e. The flux at 4.8\,GHz measured by the
RATAN-600 telescope \citep{trushkin2007} increased rapidly about 10
days after the disc became visible, i.e.\ about 5 days after the
enhanced wind and jet speeds, as depicted in
Fig.\,\ref{fig:flaresequence}f.  There is a local maximum in radio
flux at Day\,+299 and a larger flare between Day\,+300 and +302; it is
possible that these two flares could correspond to a type-1 flare
(corresponding to optically-thin bremsstrahlung radio emission from
the wind) and type-2 flare (synchrotron radio emission from expanding
jet bolides) respectively but there is insufficient spectral
information to constrain this with any certainty.  After Day\,+302 the
radio intensity falls, although there is increased radio emission
around Day\,+310.  We note that on Day +294, there are hints of a
``precursor dip" or ``quench'' in radio intensity prior to a steep
increase in intensity at the onset of a major flare, similar to the
behaviour reported by \citet{Vermeulen1993radiospectra} for the 1987
flares.  The quench phase may relate to high opacity at radio
wavelengths of the jet bolides that appear so intensely at optical
wavelengths shown in Fig.\,\ref{fig:flaresequence}d.  Such precursor
dips or quenches are also observed in the microquasar Cygnus X-3
\citep[e.g.\,][]{trushkin2007}.  In Fig.\,\ref{fig:flaresequence}c,
the intensity of H$\alpha$ radiated from the wind has a minimum just
after Day\,+294.  We do not have absolute photometry with these data
but since the instrument setup was kept the same and the weather
conditions were similar, we are confident that the intensities of the
lines during these days can be meaningfully compared with each other
and thus that this dip is a real effect.

\section{More general conclusions about flares, including those observed by others}

A priori we should expect that a flare, in the sense of enhanced
intensity observed at a particular waveband such as the radio, might
be attributable to wind or to jets and not necessarily just due to jet
ejection.  We have shown in Sec\,\ref{sec:sequence} that in a major
flare, mass-loss via both wind and jets appear to play together.  
  Collectively, the panels in Fig\,\ref{fig:flaresequence} seem to
  imply that type-1 and type-2 flares, if as we surmise correspond to
  (1) enhanced mass-loss from the wind and (2) enhanced mass-loss from
  the jets respectively, are a characteristic sequence that comprise
  this flare phenomenon.   We now consider whether the phenomena we
have observed are consistent with the characteristics of major flares
observed by others.

\subsection{Generality of wind behaviour during a flare}

\citet{Dopita1981} report that an optical outburst in SS433 is
correlated with the broadening of the stationary lines and the
appearance of P-Cygni profiles.  This is consistent with our reporting
in Sec.\,\ref{sec:optical} of the dramatic broadening of the
stationary line because of (i) the hike in wind speed and intensity
and (ii) the appearance of the widely-spaced accretion disc
lines.

\subsection{Is the disappearance of jet lines following a major flare a general characteristic of flares?}
\citet{Vermeulen1993} report a substantial reduction in the intensity
of the jet lines just as we depict in Fig.\,\ref{fig:linesdisappear}
(symmetrically in both the east and west jets as simultaneously as it
is possible to discern given the time sampling available with
  existing optical spectroscopy), within one day of an optical flare
as observed with V-band photometery by \citet{Aslanov1993}.
Similarly, \citet{Margon1984} report that particularly strong jet
lines (compared with Margon et al's $> 400$\,nights of
  observations) were followed by the disappearance of jet lines
(seemingly a ubiquitous flare characteristic) with the moving lines
returning the following month.  It is remarkable that the SS433 system
appears to return to the average behaviour predicted by the kinematic
model \citep{Margon1984} after a cessation from the usual jet
  launch following major flare activity.

\subsection{Are flares associated with elevated jet speeds?}

Although there is a tremendous archive of radio data from the Green
Bank Interferometer (GBI) spanning the time from Julian Days 2445463 to
2445509, the overlap of these data with the densely sampled
spectroscopic archive recorded by Eikenberry and Collins, as used by
\citet{BB2005} to construct a historical record of jet launch speeds,
is unfortunately rather sparse.  However, the following may be
discerned from these two sets of archival data: whenever the measured
jet-speed exceeds $0.29c$, this coincides with, or is within a day or two
of, a radio flare; there are no exceptions to this in the data
currently available.  For example, on Julian Day 2444437 the jet speed
reached 0.32$c$ and there is a clear flare recorded by the GBI where the
flux density at 2.3 GHz exceeded 1.5\,Jy (over a factor of two in
excess of the quiescent flux density at this frequency).

\citet{Margon1984} report extreme intensity variability (at least
equivalent width variability, which can be tricky to understand since
the continuum is varying) in the jet lines of SS\,433 as it undergoes
a flare in the early 1980s.  Figure\,\ref{fig:margon} shows (in the
lower panel) the speed increases at this epoch, derived using the
technique presented by \citet[][eqn\,2]{BB2005}, and (in the upper
panel) radio flux densities at 2\,GHz and 8\,GHz from the GBI.  This
figure shows the increase of the jet speeds shortly precedes a radio
flare observed at both 2\,GHz and 8\,GHz and also, as reported by
Margon et al, the temporary disappearance of the jet lines.  It is
curious that following this major flare, there are subsequent radio
flares at very similar orbital phases to the first, as observed in the
data we present in this paper, as briefly considered in
Sec\,\ref{sec:clustering}.

\subsection{Clustering, timing and phasing of flares}
\label{sec:clustering}
\begin{figure}
\begin{center}
   \includegraphics[width=8.5cm]{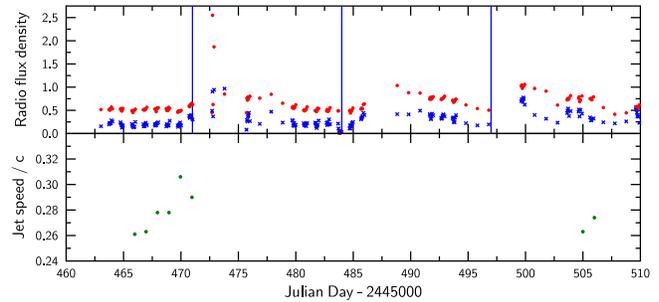} 
   \caption{Upper panel: radio flux density from the GBI, with red
     points corresponding to 2 \,GHz and blue points to 8\,GHz.  Lower
     panel: jet speeds inferred from the redshifts reported by
     \citet{Margon1984} using the technique of \citet[][eqn\,2]{BB2005}.  The
     vertical blue lines correspond to orbital phase 0.  }
\label{fig:margon}
\end{center}
\end{figure}

\citet{Feidler1987} found the time between clusters of flare peaks is
$25 \pm 10$ days while the time between adjacent clusters is $150 \pm
50$ days.  They suggested that ``the clustering of flares observed in
the radio lightcurve suggests that the reservoir of material ejected
from the binary system is quasi-periodically loaded and dumped.  The
loading and dumping timescales correspond to the quiescent period
between each cluster and the time between flare events in a cluster,
respectively."  It is not clear how to reconcile our observations of
the major flare we observed on Day\,+294 followed by its successor one
orbital period later, although we note that Fig.\,\ref{fig:margon}
seems to depict another instance of flares repeating one orbital period apart.

We note that the episode of jet flaring reported by
\citet{kopylov1985} coincided with a primary eclipse just the same
phasing as for the flare we present in this paper.
\citet{FabrikaIrsmam2003} suggest a dependence of flaring with certain
orbital phase and precession phase combinations based on a
conglomeration of radio and optical flares.  We remark that radio and
optical flares in SS433 are not going to be co-temporal because of
expansion and opacity effects, as exemplified by the sequence
represented in Fig.\,\ref{fig:timesequence}.  We also point out that
radio flaring may be due to wind as well as jets (and we note the
profile of the radio curve we show in Fig.\,\ref{fig:timesequence}f is
consistent with two peaks, resembling the case of Cygnus X-3
\citep{jcamj2011} where there is a very secure identification of a
first peak with wind ejection and a second peak with jet ejection
because of contemporaneous images made with VLBI techniques).
However, assuming flare stages are reported with accurate
time-stamping and a dependence of these on orbital phase is discerned,
we remark that this would be circumstantial evidence that there is
ellipticity in the binary orbit of SS433, supporting our discussion in
Sec\,\ref{sec:eccentricity}.

\citet{Chakrabarti2005} present a study of the variability of SS433's
emission at radio, infra-red and optical wavelengths and report that
there is typically a delay of two days between infra-red and radio
variability.  It seems plausible that this delay arises as type-2
flare peaks move from high radio frequency to low radio frequency as
bolides expand and thus transition from being optically thick at a
given frequency to optically thin.  We suggest that this time delay
relates to the time-of-flight in traversing the distance between the
nucleus of SS433 and the radio-brightening zone discussed by
\citet{Vermeulen1987} and \citet{Paragi1999}.

It is unfortunate that there are no VLBA observations of SS433
during the flare sequence that we witnessed.  However, the diagnostic
power of spectroscopic observations is exemplified throughout the
flare episode and we have been able to reconstruct the sequence of
events summarized in Figure\,\ref{fig:timesequence}. 

\section{(How) are X-rays associated with flares in SS433?}
There are famous examples of a detailed correspondence of dramatic
X-ray lightcurve variation associated with jet flaring in some
microquasars, e.g.\ GRS\,1915+105 \citep{Mirabel1998}.  In SS433, the
relation between flares and X-rays is much less clear.  Examination of
the RXTE ASM archival
database\footnote{http://xte.mit.edu/asmlc/ASM.html} for data observed
during our optical monitoring described in Sec\,\ref{sec:optical}
reveals that there is no hint of any increase in X-ray luminosity
during the flare we describe in this paper.  This lack of making any
identification of X-ray acknowledgement of the flaring activity may be
because at this epoch the jet axis is fairly close to the plane of the
sky and hence the innermost regions obscured: \citet{Nandi2005} in
fact suggest that the X-rays associated with SS433 arise from the very
inner base of the jets.  They report double normal flux rates from
RXTE when high jet speeds ($> 0.3c$) occur, possibly at epochs when
the jet axis is not aligned with the plane of the sky.  However, we
emphasize that this conclusion is contingent on Nandi et al correctly
identifying the blueshifted FeXXVI and redshifted FeXXV lines.

\section{The nature of the accretion disc during the flare}
\label{sec:lurch}

The blue accretion disc line is seen to move significantly bluewards
in Fig.\,\ref{fig:ideagram} in contrast with the more subtle bluewards
movement by the red accretion disc line over the same timerange.

One possibility is that for several days after Day\,+298 obscuring
material is progressively unveiling faster regions on the blue side of
the accretion disc while tending to cover faster regions on the red
side. Another possibility might be interference by a massive stellar
wind outflow from the companion. Extreme velocity components of
H$\alpha$ are observed intermittently in data other than ours
\citep{falamo1987,Dopita1981,kopylov1985} and the extreme excursions
on Day\,+305 (Fig.\,\ref{fig:ideagram}) are certainly real (and might
represent a brief view of regions of the disc further in).

Material from the companion star is feeding the accretion disc and may
fall towards the disc from the L1 point.  Before the
  material reaches the disc, it is possible that it is moving faster
than the orbital speed and may have to slow down via a shock thus
forming a hotspot.  If there is a trail of material approaching the
(outer) edge of the disc and glowing -- like a Katherine wheel -- then
close to phase 0 this trail could be pointing away from Earth and
could give light from a long stream or flow of material moving away
from Earth, thus influencing the wavelength at which the material in
the disc rotating away from us would be observed.
 
 \begin{figure}
    \centering
    \includegraphics[width=8.5cm]{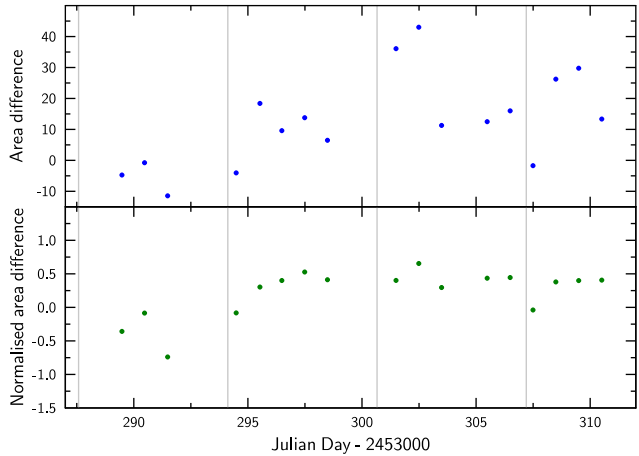} 
    \caption{{\it Upper panel:} the difference in area (i.e.\
      intensity) for the accretion disc's red and blue emission lines
      arising from H$\alpha$ emission indicating there is no orbital
      phase dependence of this quantity, in contrast to the situation
      observed in the near-IR for the Brackett-$\gamma$ lines by
      \citet{Perez2009}.  {\it Lower panel:} after Day + 294, the
      normalised difference in area appears to be independent of
      orbital phase, i.e.\ the difference in the intensities of the
      red and blue jet lines divided by their combined intensity.   }
    \label{fig:discarea}
 \end{figure}
   
 There is no evidence of the sinusoidal variation in the difference of
 the areas of the Balmer H$\alpha$ accretion disc lines, in comparison
 with the clear sinusoid seen for the areas of the
 Brackett-$\gamma$ lines \citep{Perez2009}.  Fig.\,\ref{fig:discarea}
 shows (upper panel) the complete lack of any sinusoidal dependence
 while the lower panel reveals a rather steady normalised difference
 in line areas (the difference in the intensities of the red and blue
 jet lines divided by their combined intensities) which indicates no
 dependence on orbital phase at all, during the interval the accretion
 disc is revealed by optical spectroscopy.  This implies that
 the stellar radius is rather smaller after the flare was instigated
 (since it has no attenatuating role like that observed in the
 near-IR by \citet{Perez2009}) and so may indicate that a considerable
 wind was blown off the star prior to the build-up of the flare
 itself.

\section{Conclusions}

This sequence of observations has, for the first time in the optical,
very clearly identified the accretion disc of SS433, although it seems
likely that \citet{falamo1987} had a glimpse in observations made in
June 1979.  Subsequently, both \citet{Dopita1981} and
\citet{kopylov1985} made separate observations of the same
flare episode in July 1980.  More recently \citet{Perez2010} have
presented evidence for the accretion disc lines viewed in the optical
through the attenuating disc wind while \citet{Perez2009} presented
infra-red spectroscopy of the of the accretion disc.  The centres of
both the accretion disc and the wind during the flare episode are
orbiting at speeds consistent with 175\,km\,s$^{-1}$ --
200\,km\,s$^{-1}$, in agreement with the He\,II and C\,II measures of
the orbital speed of the compact object about the centre of mass of
the system \citep{Crampton1981,Fabrika1990}.  We suggest the flare was
initiated by a disturbance of the outer regions of the accretion disc,
perhaps a massive ejection from the companion which fed the accretion
disc. After some delay the disc responded with a doubling of wind
speed, initially reduced jet speed then dominated by a period of higher
than usual launch speeds and finally a complex radio flare.  
  Collectively, the panels in Fig\,\ref{fig:flaresequence} seem to
  imply that type-1 and type-2 flares, if corresponding to (1)
  enhanced mass-loss from the wind and (2) enhanced mass-loss from the
  jets respectively, are a characteristic sequence that comprise the
  flare phenomenon. 

\section*{Acknowledgements}
The new observations reported in this paper were made possible by the
grant of Director's Discretionary Time on the 3.5-m New Technology
Telescope. We thank Jochen Liske for his excellent line-fitting
software, SPLOT.  We acknowledge the quick-look X-ray monitoring
provided by the ASM/RXTE team.  It is a pleasure to thank Stephen
Blundell, Stephen Justham, Michael Bowler, Sebastian Perez and Samuel
Doolin for encouragement and discussions.  KMB is very grateful to the
Royal Society for a University Research Fellowship.  ST thanks the
Russian Foundation for Basic Research for support, grant
N\,08-02-00504-a.    We warmly thank the referee for careful comments
on the manuscript. 

\bibliographystyle{mn2e} %% MNRAS

\end{document}